\documentclass[a4paper,fleqn]{cas-dc}

\usepackage[utf8]{inputenc}   % harmless on pdfLaTeX; drop for LuaLaTeX/XeLaTeX
\usepackage{textcomp}

\usepackage[numbers,sort&compress]{natbib}

\usepackage[acronym,abbreviations]{glossaries-extra}
\setabbreviationstyle[acronym]{long-short}

\newacronym{minlp}{MINLP}{Mixed-Integer Non-Linear Program}
\newacronym{miqp}{MIQP}{Mixed-Integer Quadratic Program}
\newacronym{socp}{SOCP}{Second Order Cone Program}
\newacronym{dnr}{DNR}{Distribution network reconfiguration}
\newacronym{npf}{NPF}{number of power-flow evaluations}

\graphicspath{{images/}}
\DeclareGraphicsExtensions{.pdf,.png,.jpeg,.jpg}

\usepackage{subcaption}
\usepackage{listings}
\hypersetup{
    linkcolor=black,
    citecolor=blue,
    urlcolor=blue,
    pdftitle={Distribution Network Reconfiguration: A Reproducible Benchmark of
              Heuristic, Metaheuristic and Mathematical Methods}
}

\newenvironment{cellitemize}
  {\begin{list}{\labelitemi}{%
      \setlength{\labelwidth}{0.5em}%
      \setlength{\labelsep}{0.35em}%
      \setlength{\leftmargin}{0.85em}%
      \setlength{\rightmargin}{0pt}%
      \setlength{\itemindent}{0pt}%
      \setlength{\listparindent}{0pt}%
      \setlength{\topsep}{2pt}%
      \setlength{\partopsep}{0pt}%
      \setlength{\parsep}{0pt}%
      \setlength{\itemsep}{2pt}%
   }}
  {\end{list}}

\begin{document}

\let\WriteBookmarks\relax
\def\floatpagepagefraction{1}
\def\textpagefraction{.001}

% -----------------------------------------------------------------------------
%  FRONT MATTER
% -----------------------------------------------------------------------------

\shorttitle{Distribution network reconfiguration: a reproducible benchmark}
\shortauthors{F. Bohigas-Daranas et al.}

\title[mode=title]{Distribution Network Reconfiguration: A Reproducible
Benchmark of Heuristic, Metaheuristic and Mathematical Methods}

\tnotemark[1]
\tnotetext[1]{This work was supported by project TED2021-130351B-C21 (HP2C-DT),
funded by MICIU/AEI/10.13039/501100011033 and by the European Union
NextGenerationEU/PRTR, and by the Daedalos project.}

% Add ORCIDs when available, e.g.
\author[1]{Ferran Bohigas-Daranas}[orcid=0009-0001-6477-6591]
\cormark[1]
\ead{ferran.bohigas@upc.edu}

\affiliation[1]{organization={CITCEA, Universitat Polit\`ecnica de Catalunya},
            addressline={Av.\ Diagonal 647},
            city={Barcelona},
            postcode={08028},
            country={Spain}}

\author[1]{Eduardo Prieto-Araujo}
% \credit{Supervision, Writing -- review \& editing}

\author[1]{Oriol Gomis-Bellmunt}
% \credit{Supervision, Funding acquisition}

\cortext[1]{Corresponding author}

% -----------------------------------------------------------------------------
%  ABSTRACT
% -----------------------------------------------------------------------------
\begin{abstract}
Distribution network reconfiguration (DNR) has been studied for five decades, yet
published algorithms are rarely compared under identical assumptions: authors report
results on different test systems, with different power-flow models, and almost never
release code. This paper re-implements eight representative DNR algorithms in a
single open-source Python framework and benchmarks them on five networks under
identical conditions. The algorithms span three of the four established paradigms: heuristics
(loop cutting, branch exchange, greedy minimum spanning tree, exhaustive search),
metaheuristics (a genetic algorithm and a selective binary particle swarm optimizer) and mathematical programming (a mixed-integer quadratic program built on a convex, simplified-DistFlow power-flow model). We report power losses, voltage profile, and, as a hardware-independent cost metric, the number of power-flow evaluations required for convergence.
On the 196-bus Simbench urban network, we show that a dynamic DNR policy can save more than 26\% of the system losses, weighed against circuit-breaker maintenance cost. All implementations, test cases, and result scripts
are released to allow direct extension and replication.
\end{abstract}

% -----------------------------------------------------------------------------
%  HIGHLIGHTS  (required by most Elsevier journals; derived from the
% -----------------------------------------------------------------------------
\begin{highlights}
\item Eight DNR algorithms re-implemented in one open-source Python framework.
\item All methods benchmarked on five networks under identical conditions.
\item Power-flow evaluation count used as a hardware-independent cost metric.
\item A year-long dynamic policy recovers more than 26\% of the network losses.
\item Switching cost is weighed against energy savings via breaker endurance.
\end{highlights}

% -----------------------------------------------------------------------------
%  KEYWORDS
% -----------------------------------------------------------------------------
\begin{keywords}
Distribution network reconfiguration \sep
heuristic algorithms \sep
minimum spanning tree \sep
metaheuristic algorithms \sep
genetic algorithm \sep
particle swarm optimization \sep
mixed-integer quadratic programming \sep
benchmarking \sep
open-source software
\end{keywords}

\maketitle

\section{Introduction}
\label{introduction}

Distribution networks can be built with meshed topologies but are almost always
operated radially, because radial operation simplifies protection coordination and
limits short-circuit currents. Radiality is not free, it forces power to travel along
a single path to each load, and the resulting resistive losses account for the
majority of the energy lost between generation and consumption. Total network losses
reach up to 15\% of the transferred energy worldwide and average 8\% in Europe\cite{CEER_2nd_nodate}. In
Spain, with a yearly electricity consumption close to 250 TWh, they amount to roughly 20\,TWh per year, which, at average wholesale prices, are worth close to 1.6 billion euros \cite{omie_2025}.

\gls{dnr} reduces these losses without new infrastructure. By opening normally-closed sectionalizing switches and closing normally-open tie switches, the operator changes which radial tree the network forms, and therefore how current is routed. It is one of several levers a utility can pull to cut losses, and the only one that adds no new hardware—it simply re-routes power through switches that are already in place; Section \ref{sec:other_methods} contrasts it with the capital-intensive alternatives. The growth of distributed energy resources, whose output varies on a timescale of minutes, has turned DNR from a seasonal planning exercise into a candidate for near-real-time operation, enabled by SCADA and remote switch control.

The obstacle is computational. A network with $N_s$ switchable lines admits $2^{N_s}$
candidate configurations, of which only the spanning trees are feasible. Expressed as
an optimization problem, DNR couples binary switch states to the nonconvex power flow
equations, producing a \gls{minlp}; even after the standard conic
relaxation of the branch flow equations, the binary variables leave it NP-hard
\cite{taylor_convex_2012}. No polynomial-time algorithm is available, and the
literature has therefore split into three solution families---heuristics, metaheuristics, and
mathematical programming---each trading optimality against tractability differently.

Several surveys have mapped this literature \cite{tang_survey_2014,
kalambe_loss_2014, milad_rahimipour_comprehensive_2023, mahdavi_reconfiguration_2021}.
They share a limitation: they compare methods using the results each original author
reported. Those results were obtained on different test systems, with different
power-flow models, different convergence tolerances, and different hardware, and the
underlying code is almost never available. A reader cannot therefore tell whether one
algorithm beats another because of its search strategy or because of its power-flow
approximation. Reported execution times compound the problem, since they age with the
hardware they were measured on \cite{pereira_distribution_2023}.

This paper takes the opposite approach. Rather than tabulate published numbers, we
re-implement eight representative algorithms from scratch in a single Python
framework, drive them all with the same power-flow engine and the same convergence
criteria, and run them on the same five networks. Where an algorithm's original
formulation is ambiguous, we state the interpretation we adopted. All code is
released.

The contributions are:
\begin{enumerate}
    \item An open-source Python library \cite{dnr_github_2025} containing all eight implementations, the five test cases, and the scripts that generate every figure and table in this paper.

    \item A controlled comparison of the algorithms on identical networks, reporting losses, voltage profile, and the \gls{npf}. We argue for \gls{npf} over time as the primary cost metric, since it is invariant to hardware, and we quantify the per-evaluation cost of each power-flow model so that \gls{npf} can be converted back to time. Only in the case of the mathematical based algorithm, the use of \gls{npf} loses its sense, as it solves the power flow equations internally.

    \item An evaluation over a full year of hourly load and generation data on a 196-bus network, quantifying the energy actually recoverable by a dynamic DNR policy and the number of switching operations it requires.

    \item The metric definition to study the circuit breaker aging caused by the dynamic \gls{dnr} against the energy savings, to analyze the economical feasibility of the dynamic DNR implementation.
\end{enumerate}

Section~\ref{sec:the_problem} states the problem and its constraints, including the power-flow models and the radiality condition. Section~\ref{sec:mathematical-formulation-and-constraints} presents the mathematical formulation and constraints. Section~\ref{Methods-Review}
classifies and describes the methods.
Section~\ref{sec:results} reports and discusses the results. Section~\ref{sec:software} presents the software.
Section~\ref{sec:conclusion} concludes.

\section{Understanding the problem}
\label{sec:the_problem}

\subsection{Distribution Network Reconfiguration Objectives}
\label{sec:dnr_fundamentals}

\gls{dnr} aims to optimize system operation for enhanced reliability and economy, primarily addressing four objectives: Power Loss Reduction, Voltage Profile Optimization, Load Balancing, and Service Restoration. Of the eight methods implemented, seven were initially dedicated to Power Loss Reduction, with only one's initial formulation incorporating Voltage Profile as main objective \cite{salkuti_effective_2021}. In contrast, our implementation has expanded these metaheuristic methods to be multi-objective, thereby enabling the optimization of both power losses and voltage profile.

\subsubsection{Power Loss Reduction}
Minimizes resistive losses ($P_{loss}$) through optimal path selection:
\begin{equation}\label{eq:loss}
\min \sum_{k=1}^{N_l} I_k^2 R_k \quad \forall k \in \{1,\dots,N_l\}
\end{equation}
where $I_k$ is the current on line k, which is constrained by its line ampacity $I_k \leq I_{k}^{max}$, $R_k$ is the resistance of that line, and $N_l$ is the number of lines in the electrical network.

\subsubsection{Voltage Profile}
A significant effect of the radial topology is increased voltage loss due to voltage drop in the cables, so maintaining the voltages within the DSO's limits can be another objective in the optimization process:
\begin{equation}\label{eq:voltage_profile}
V_i \in [V_{min}, V_{max}] \quad\forall i \in \{1,\dots,N_b\}
\end{equation}
where $V_i$ is the voltage at bus i measured in pu, $N_b$ is the number of buses in the electrical network, while $V_{min}$ and $V_{max}$ are the maximum and minimum voltages defined by the DSO.

\subsection{Solution Strategies for Distribution Optimization}\label{sec:other_methods}

\gls{dnr} is not the only means available to a utility for reducing losses and improving voltage profiles. Shunt-capacitor placement and distributed-generation (DG) siting pursue the same goals during the planning stage, but both add physical assets and therefore fall on the capital-expenditure (CAPEX) side, whereas DNR re-uses installed switches and is an operational-expenditure (OPEX) measure. The three are compared below.

\subsubsection{Shunt Capacitor Placement}
This approach \cite{kalambe_loss_2014} improves systems by compensating reactive power and stabilizing voltage. This method requires installing fixed capacitors at specific locations. Benefits include reduced losses and better voltage profiles, though these goals may sometimes conflict. Limitations include high upfront costs, static nature, need for placement analysis, and large incremental adjustments.

\subsubsection{DG Allocation}

DG Allocation \cite{kalambe_loss_2014} strategically positions distributed generation sources close to consumers, which minimizes energy transmission distances and consequently reduces power losses. Furthermore, DGs enhance voltage profiles by increasing voltage levels at the far ends of the network and alleviating cable loading. However, the integration of DGs can introduce potential challenges related to reverse power flow. Similar to Shunt Capacitor Placement, implementing DG Allocation necessitates substantial financial investment and represents a static infrastructure solution.

\subsubsection{Distribution Network Reconfiguration (DNR)}
It alters network topology via switch exchanges, offering flexibility at minimal cost. It reduces losses, improves voltage profiles, and isolates faults. The challenge lies in finding optimal configurations, a non-convex, discrete problem that conventional methods cannot easily solve — the subject of the remainder of this paper.

The authors are aware of the limited life of switches when they operate under load. Future work will investigate adapting the base strategies to account for the available electrical elements, as mentioned in Section \ref{sec:conclusion}.

\subsection{Intuition behind the DNR methods}\label{DNR-intuition}

\begin{figure*}[width=\textwidth,pos=tbp]
    \centering
    \begin{subfigure}[t]{0.32\linewidth}
        \centering
        \includegraphics[width=\linewidth]{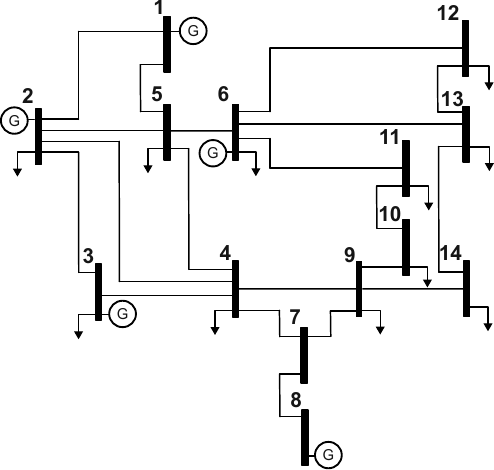}
        \caption{IEEE 14-bus case, with all switches closed and losses equal to 13.34 MW.}
        \label{fig:14buses}
    \end{subfigure}
    \hfill
    \begin{subfigure}[t]{0.32\linewidth}
        \centering
        \includegraphics[width=\linewidth]{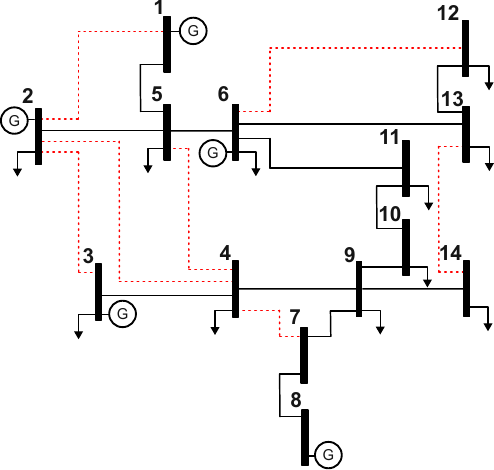}
        \caption{Non-optimal configuration with losses equal to 115.23 MW.}
        \label{fig:figure1}
    \end{subfigure}
    \hfill
    \begin{subfigure}[t]{0.32\linewidth}
        \centering
        \includegraphics[width=\linewidth]{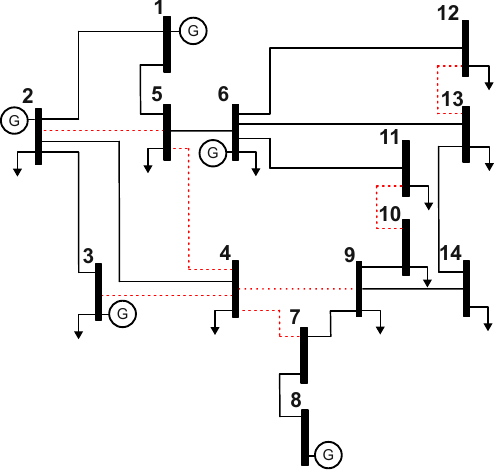}
        \caption{Near-optimal configuration with losses equal to 28.91 MW.}
        \label{fig:figure2}
    \end{subfigure}
    \caption{IEEE 14-bus case non-optimal and near-optimal configurations, where disabled lines are represented in red.}
    \label{fig:14buses_optimal_non_optimal}
\end{figure*}

To clearly exemplify the DNR problem, we will use the IEEE 14-bus system, shown in Fig. \ref{fig:14buses}, a transmission network that has been used as DNR example, due to its size and simplicity, despite the power values. This network comprises 14 buses, 20 lines, 5 generators, and 11 loads, and includes 7 loops. The objective of DNR methods is to achieve a radial architecture by opening 7 switches to eliminate the loops, while keeping all nodes energized, selecting the switches to open or close, such that the total network losses are minimized.

We will denote the line names as follows: the line connecting buses 6 and 12 is named '6\_12\_1'. If multiple lines connect the same two buses, the trailing number will enumerate them.

If we choose a random radial configuration, defined by its disconnected lines, Fig.~\ref{fig:figure1}, such as  \textit{['6\_12\_1', '13\_14\_1', '2\_4\_1','4\_5\_1','4\_7\_1','2\_3\_1','1\_2\_1']}, the losses will be 115.23 MW with a voltage range between 0.815 and 1.09 p.u.

Executing the method formulated by \cite{merlin_search_1975}, Fig.~\ref{fig:figure2}, chooses the lines \textit{['2\_5\_1', '3\_4\_1', '4\_5\_1', '4\_7\_1', '4\_9\_1', '10\_11\_1', '12\_13\_1']}. This solution achieves 28.91 MW of losses, maintains network radiality, and improves the voltage profile to a range of 0.958 to 1.09 p.u., resulting in lower losses and enhanced network parameters.

\section{Mathematical Formulation and Constraints}\label{sec:mathematical-formulation-and-constraints}

\gls{dnr} can be formulated as a traditional optimization problem requiring the definition of specific terms. Subsequently, we will outline the optimization equations and discuss the various approaches employed by authors to calculate network losses using different power flow calculation methods.

\subsection{Optimization Framework}
The DNR problem is defined as:
\begin{align}\label{eq:opt}
\min_x \quad & f(x) \quad \text{(objective function, i.e. network losses)}\\
\text{s.t.} \quad & g(x) = 0 \quad \text{(Power flow equations)} \label{eq:equal}\\
& h(x) \leq 0 \quad \text{(Operational limits)} \label{eq:inequal}\\
& x \in \{0,1\}^{N_s} \quad \text{(Switch states)} \label{eq:binary}
\end{align}
\subsection{Power Flow Models}\label{sec:powerflow}

The selection of the equations to be used in the optimization objective function (\ref{eq:opt}) and constraints (\ref{eq:equal},\ref{eq:inequal},\ref{eq:binary}) is a key decision in the solution process, as different authors have chosen different methods for solving the power flow equations (\ref{eq:equal}), affecting both the resolution time and accuracy:

\subsubsection{Power Flow Equations}

The power flow equations, defined by Kirchhoff's laws \cite{Kirchhoff1845}, are applicable to any network regardless of its topology:

\begin{align}
P_i &= \sum_k |V_i||V_k|(G_{ik}\cos\theta_{ik} + B_{ik}\sin\theta_{ik}) \label{eq:pf_p}\\
Q_i &= \sum_k |V_i||V_k|(G_{ik}\sin\theta_{ik} - B_{ik}\cos\theta_{ik}) \label{eq:pf_q}
\end{align}
where \(P_i\) is the active power flow between buses \(i\) and \(k\), \(Q_i\) is the reactive power flow between buses \(i\) and \(k\), \(V_i\) is the voltage at bus \(i\), \(V_k\) is the voltage at bus \(k\), \(G_{ik}\) is the conductance of the line between buses \(i\) and \(k\), \(B_{ik}\) is the susceptance of the line between buses \(i\) and \(k\), \(\theta_{ik}\) is the phase angle difference between buses \(i\) and \(k\).

They can be solved by methods like Newton-Raphson \cite{Raphson_analysis_1697}, among others. However, authors have applied a variety of relaxations in order to obtain solutions for the DNR problem \cite{baran_network_1989,jabr_minimum_2012,llorens-iborra_mixed-integer_2012}, as explained in Section \ref{mathematical-techniques} and Section \ref{heuristic-techniques}.

\subsubsection{DistFlow Equations}

DistFlow equations \cite{Baran1989} are a set of equations based on Kirchhoff's laws, specifically tailored to the characteristics of distribution grids, such as their radial or weakly meshed topology. They can be solved using the recursive Forward (\ref{eq:distflow_f_p1}-\ref{eq:distflow_f_v})/Backward (\ref{eq:distflow_b_p1}-\ref{eq:distflow_b_q2}) sweep method, which allows for a non-derivative solution and faster computation.

\begin{equation}
P_{i+1} = P_i - r_i \frac{P_i^2+Q_i^2 }{V_i^2} - P_{L{i+1}} \quad \label{eq:distflow_f_p1}
\end{equation}
\begin{equation}
Q_{i+1} = Q_i - x_i \frac{P_i^2+Q_i^2 }{V_i^2} - Q_{L{i+1}} \quad\label{eq:distflow_f_q1}
\end{equation}
\begin{equation}
V_{i+1}^2 = V_i^2 -  2(r_iP_i + x_iQ_i)+(r_i^2+x_i^2)\frac{P_i^2+Q_i^2 }{V_i^2} \quad\label{eq:distflow_f_v}
\end{equation}

\begin{equation}
P_{i-1} = P_i + r_i \frac{P_i^{'2}+Q_i^{'2} }{V_i^2} + P_{Li} \quad \label{eq:distflow_b_p1}
\end{equation}
\begin{equation}
Q_{i-1} = Q_i + x_i \frac{P_i^{'2}+Q_i^{'2} }{V_i^2} + Q_{Li} \quad\label{eq:distflow_b_q1}
\end{equation}
\begin{equation}
V_{i-1}^2 = V_i^2 +  2(r_iP^{'}_i + x_iQ_i^{'})+(r_i^2+x_i^2)\frac{P_i^{'2}+Q_i^{'2} }{V_i^2} \quad\label{eq:distflow_b_v}
\end{equation}
\begin{equation}
P_i^{'} = P_i + P_{Li}\quad\label{eq:distflow_b_p2}
\end{equation}
\begin{equation}
Q_i^{'} = Q_i + Q_{Li}\quad\label{eq:distflow_b_q2}
\end{equation}

\noindent where \(P_i\) is the active power flow between buses \(i\) and \(i+1\), \(Q_i\) is the reactive power flow between buses \(i\) and \(i+1\), \(P_{Li}\) is the active power flow for the loads at bus \(i\), \(Q_{Li}\) is the reactive power for the loads at bus \(i\), \(V_i\) is the voltage at bus \(i\), \(r_{i}\) is the resistance of the line between buses \(i\) and \(i+1\), \(x_i\) is the reactance of the line between buses \(i\) and \(i+1\).

During the backward sweep, powers/currents are aggregated node by node from the network extremities to the substation, while during the forward sweep, voltages are calculated from the slack bus at the main substation to the feeder extremities.

A simplified version was formulated by \cite{baran_network_1989} to achieve even faster solutions, by not considering the quadratic terms of the losses during the forward (\ref{eq:distflow_f_p1})(\ref{eq:distflow_f_q1}) and backward passes (\ref{eq:distflow_b_p1})(\ref{eq:distflow_b_q1}).

\subsubsection{Direct Load Flow (DLF)}
The method proposed by \cite{jen-hao_teng_direct_2003}, is based on the bus-injection to branch-current (BIBC) and branch-current to bus-voltage (BCBV) matrices. It is solved by a fast recursive process, achieving a non-derivative solution simply by multiplying matrices based on the admittances and the network's interconnection:
\begin{equation}
[I_{branch}] = [BIBC][I] \quad \label{eq:dlf_b}
\end{equation}
\begin{equation}
[\Delta V] = [BCBV][BIBC][I] \label{eq:dlf_v}
\end{equation}

\noindent where \([I]\) is the vector of bus current injections, \([I_{branch}]\)
is the vector of branch currents, and \([\Delta V]\) is the vector of bus voltage
drops with respect to the substation bus.

Power flow solutions present a trade-off between accuracy and computational expense. Power flow equations (\ref{eq:pf_p},\ref{eq:pf_q}) when solved by Newton-Raphson method achieve the highest accuracy but at the highest computational cost. In contrast, DistFlow and Direct Load Flow (DLF) equations offer the fastest solution at the expense of only being suitable for radial or weakly meshed networks.

Considering that solving the power flow equations is the most time-consuming task within network reconfiguration methods, as this operation can be done iteratively for tens or up to thousands of times, the choice of how to solve these equations directly impacts the solution time. For this reason, some authors choose to apply certain relaxations, assumptions, or simplifications to the equations to improve the power flow solution time. The most common approaches include:
\begin{itemize}
    \item Assuming cable losses are negligible compared to loads, allowing them to be omitted from the equations.
    \item Setting voltages in all nodes to 1 per unit (pu).
    \item Considering only the real part of the equations (as commonly done in older works).
\end{itemize}

In heuristic methods, the equations are manipulated as part of the decision-making process \cite{baran_network_1989,civanlar_distribution_1988}, whereas metaheuristics typically use an external power flow solver to obtain the objective function result. Mathematical approaches \cite{jabr_minimum_2012,llorens-iborra_mixed-integer_2012} utilize these power flow equations as constraints or objectives, as detailed in Section \ref{mathematical-techniques}.

Each author selects the objective function based on their decision criteria, with the most common being the minimization of power losses in the network, but other criteria could include voltage margins, switching costs, or a combination of weighted objectives.

\subsection{Network radiality}

Special attention has been given to the definition of radiality in the network and its translation into mathematical equations that can be easily incorporated into the decision-making process.

A radial network requires that there are no loops in the topology, which can be mathematically defined as:

\begin{equation}
N_{l}\; = \; N_{b} - N_{con} \label{eq:radiality_1}
\end{equation}
\noindent where \(N_{l}\) is the number of available lines, \(N_{b}\) is the number of buses and \(N_{con}\) is the number of connections between the network and the main transmission or subtransmission grid. In the cases shown in this paper, the number of connections \(N_{con}=1\).

However, this condition alone is not sufficient; it also requires that all buses are energized.

In heuristic and metaheuristic approaches, the radiality constraint is verified by equation \eqref{eq:radiality_1}, while ensuring that no nodes are disconnected or de-energized after each decision or selection step. In contrast, mathematical optimization methods require the radiality constraint to be integrated into a mathematical equation, as defined by \cite{jabr_minimum_2012} and \cite{lavorato_imposing_2012}.

The necessary and sufficient conditions for radiality in a graph, and thus for the existence of a spanning tree, are rigorously established through the calculation of the determinant of the branch-to-node incidence matrix \cite{abur_modified_1996}. Nevertheless, mathematical optimization techniques are generally incapable of incorporating determinants directly as constraints, thus necessitating their expression through alternative mathematical formulations. One of the main contributions to this topic is presented in \cite{lavorato_imposing_2012}, where the authors define radiality as satisfying two simultaneous conditions:

\begin{itemize}
    \item Condition 1: The network configuration must form no loops.
    \item Condition 2: The network configuration must have all buses connected.
\end{itemize}

In \cite{jabr_minimum_2012} and \cite{gallego_mixed-integer_2022}, the authors introduce two binary
variables, $z_{ij}$ and $z_{ji}$, to indicate the power-flow direction along
a line. Together with the line's switch status $y_{ij}$, they satisfy

\begin{equation}
    z_{ij} + z_{ji} = y_{ij}, \qquad y_{ij} \in \{0,1\},
    \label{eq:direction}
\end{equation}

where $z_{ij} = 1$ denotes flow from bus $i$ to bus $j$. When the line is in
service ($y_{ij} = 1$), exactly one of the two directions is active; when the
line is open ($y_{ij} = 0$), both $z_{ij} = z_{ji} = 0$ and the branch carries
no current, so it is effectively disconnected.

\subsection{Graph Theory}
Electrical networks can be modeled as undirected graphs, \textit{G(V,E)}, where nodes correspond to network buses, and edges represent the lines connecting them \cite{morton_efficient_2000}. This representation allows graph optimization methods and algorithms to be readily applied to electrical systems. For the \gls{dnr} problem, the most suitable algorithms are those designed to find the Minimum Spanning Tree (MST)\cite{montoya_minimal_2012}, such as Kruskal's \cite{Kruskal1956} or Prim's \cite{Prim1957} algorithms. These ensure the radial structure of the distribution system while optimizing the paths based on various criteria, such as line impedance or the power flowing through the lines, which are used as edge weights.

\section{Methods Review}\label{Methods-Review}

\subsection{Solution Approaches Classification}
\begin{figure}[pos=htbp]
    \centering
    \includegraphics[width=0.5\linewidth]{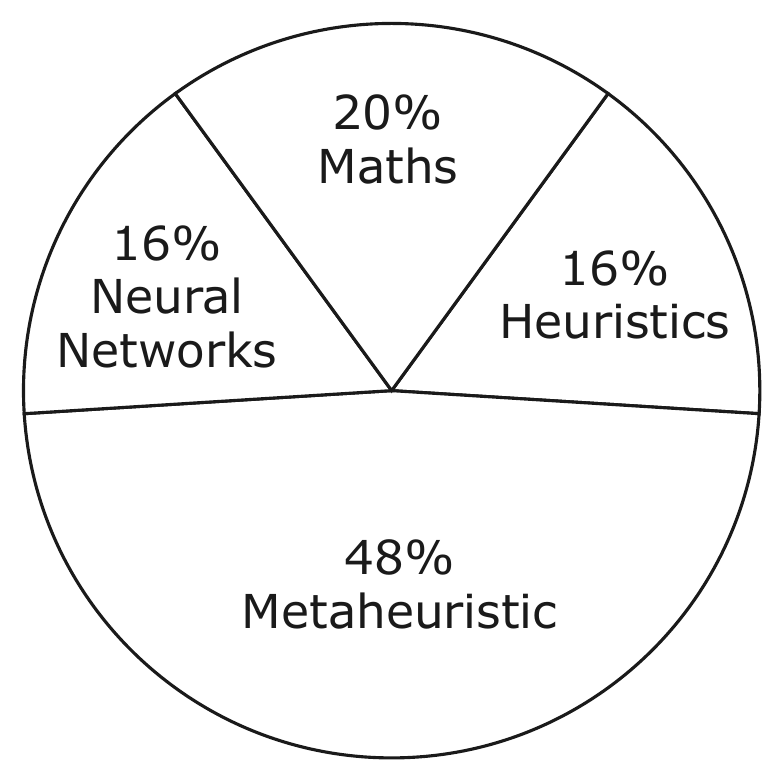}
    \caption{Percentage of the different DNR methods on the technical literature, based on the information given in \cite{mahdavi_reconfiguration_2021} and other papers. The \emph{Neural Networks} share corresponds to the data-driven family of Table~\ref{tab:methods_summary}.}
    \label{fig_methods_percentage}
\end{figure}

The evolution of DNR methodologies can be categorized into four distinct paradigms, summarized in Table \ref{tab:methods_summary}. Fig.~\ref{fig_methods_percentage} shows their relative share of the published literature.

Of these four, the first three, mathematical programming, heuristics, and metaheuristics, are reimplemented and benchmarked in this paper (Sections \ref{mathematical-techniques}–\ref{metaheuristic} and Section \ref{sec:results}). The fourth, data-driven methods, is included in the classification for completeness but is deliberately left outside the benchmark. Unlike the other three, a learned model does not solve an instance from its physics; it must first be trained on a representative corpus of already-solved instances (supervised learning) or through repeated interaction with a simulated network (reinforcement learning), and its solution quality is then a function of that training set and of generalization to unseen topologies and load patterns. A fair evaluation would therefore measure a fundamentally different quantity than the one this study isolates, the effect of search strategy and power-flow relaxation under identical conditions, and would require its own training protocol and out-of-distribution test design. We accordingly review the data-driven family here and defer its benchmarking to future work (Section \ref{sec:conclusion}). The remainder of this section describes the three benchmarked families in turn.

\begin{table*}[width=\FullWidth,pos=tbp]
    \caption{\gls{dnr} methods summary}
    \label{tab:methods_summary}
    \centering
    \footnotesize
    \begin{tabular}{|>{\centering\arraybackslash}m{0.11\linewidth}|>{\raggedright\arraybackslash}m{0.19\linewidth}|>{\raggedright\arraybackslash}m{0.23\linewidth}|>{\raggedright\arraybackslash}m{0.28\linewidth}|>{\raggedright\arraybackslash}m{0.07\linewidth}|} \hline
         \textbf{Heuristic Methods}&
         The foundational methods in DNR employ rule-based approaches that utilize network physics and practical rules derived from network-specific knowledge &
        \begin{cellitemize}
            \item Branch exchange algorithms
            \item Loop-based algorithms
            \item Minimum Spanning Tree based methods
        \end{cellitemize} &
        \begin{cellitemize}
            \item Fast computation (polynomial time complexity)
            \item Intuitive physical interpretation
            \item Local optima convergence
            \item Dependency on initial conditions
        \end{cellitemize}&
            \cite{baran_network_1989,merlin_search_1975,civanlar_distribution_1988,montoya_minimal_2012}
        \\ \hline

        \textbf{Metaheuristic Methods}&
        Nature-inspired optimization techniques, which include genetic and swarm intelligence approaches     &
        \begin{cellitemize}
            \item Genetic Algorithms (GA)
            \item Particle Swarm Optimization (PSO)
            \item Ant Colony Optimization (ACO)
            \item Simulated Annealing (SA)
        \end{cellitemize}&
        \begin{cellitemize}
            \item No guarantee of global optimality
            \item Fitness evaluation via power flow solutions
            \item Population-based search
        \end{cellitemize}&
            \cite{mahdavi_reconfiguration_2021,jakus_optimal_2020,khalil_reconfiguration_nodate}
        \\ \hline

         \textbf{Mathematical Programming} & Exact optimization methods providing rigorous solutions &
        \begin{cellitemize}
            \item Mixed-Integer Linear Programming (MILP)
            \item Mixed-integer quadratic programming (MIQP)
            \item Second-order cone programming (SOCP)
        \end{cellitemize}&
        \begin{cellitemize}
            \item NP-hard problem complexity
            \item Approximation/relaxation techniques often required
            \item Incorporation of radiality constraints (Eq. \ref{eq:radiality_1})
        \end{cellitemize}&
            \cite{jabr_minimum_2012,taylor_convex_2012,llorens-iborra_mixed-integer_2012}
        \\ \hline

         \textbf{Data-driven algorithms} & Emerging data-driven techniques that have been gaining prominence in the last few years \cite{mahdavi_reconfiguration_2021} &
        \begin{cellitemize}
            \item Supervised and unsupervised neural networks
            \item Reinforcement learning for dynamic reconfiguration
        \end{cellitemize}   &
        \begin{cellitemize}
            \item Increased computational cost during the training stage
            \item Extremely fast solution in run time operation
            \item Capability to learn and generalize complex non-linear relationships
            \item Black-box nature with limited explainability
        \end{cellitemize}  &
            \cite{Meinecke_Sarajlic_Drauz-Mauel_Klettke_Lauven_Rehtanz_Moser_Braun_2020,haider_siphyr_2024}
        \\  \hline
    \end{tabular}
\end{table*}

\subsection{Mathematical techniques}\label{mathematical-techniques}

The most straightforward approach to solving the DNR problem would be to use mathematical methods based on the power flow equations. However, its mixed-integer nonlinear nature, which renders it a highly combinatorial problem, makes its solution a challenging task.

Mathematical methods possess a key advantage compared to heuristic and metaheuristic techniques (which will be discussed in Sections \ref{heuristic-techniques} and \ref{metaheuristic}): they can obtain globally optimal solutions if the problem can be transformed into a convex problem through relaxations. Moreover, they do not require any tuning or random initialization that could lead to suboptimal or inconsistent solutions.

One of the earliest references for solving the DNR problem using mathematical optimization processes is presented in \cite{abur_modified_1996}, published in 1996, which employed linear programming. The formulation is given by

\begin{align}
[A][P_f] = [P] \label{eq:opt1}
\end{align}

\noindent where \(A\) is the node-to-branch reduced incidence matrix, \(P_f\) is the branch flow vector, and [P] is the nodal injection vector, excluding the reference node.

If there are \(N_b\) buses, only \(N_b-1\) out of \(N_l\) lines need to be connected (\ref{eq:radiality_1}). However, the [A] matrix will change for each choice of \(N_b-1\) lines while maintaining radiality and connectivity. The article establishes resistive losses as the objective function, focusing solely on real power. Its main challenge lies in maintaining radiality, achieved through the use of an extended matrix.

Instead of purely linear programming, other approaches such as \cite{han_branch_2002}, which focus on communication networks but propose theoretical foundations applicable to distribution networks, suggest using the branch-and-bound method. This search method is adapted to the binary nature of communication networks and can also be applied to energy distribution networks, emphasizing efficient pruning of all possible configurations.

The DNR problem can utilize methods developed for Optimal Power Flow (OPF), a well-established field aiming to minimize one or more objective functions, typically to minimize the generation costs. However, in the case of the DNR, power losses are used as objective function. However, the added complexity lies in determining the status of switches, represented as binary variables, which makes it a \gls{minlp} problem.

Many authors focus on applying relaxations to the quadratic equations of power flow, converting them into linear equations suitable linear programming solvers. This includes Mixed-Integer Linear Programming (MILP), which integrates power flow equations and network operational constraints.

Two prominent works in the field, \cite{borghetti_mixed-integer_2012} and \cite{ramos_path-based_2005}, propose different approaches using Piecewise Linear Functions (PLF) to linearize the quadratic equations that relate power, voltage and current. This linearization process, however, adds complexity to the solution when combined with binary variables defining switch status. PLFs can effectively linearize these equations, enabling their use in MILP formulations.

In \cite{borghetti_mixed-integer_2012}, the author introduces a novel method for modeling switches based on the definition of all possible paths between nodes and substations. Each path is mathematically represented by a binary variable, ensuring radiality while simultaneously defining switch statuses. In later papers, the process has been refined and expanded to more sophisticated methods, such as mixed-integer conic programming \cite{jabr_minimum_2012}. Nevertheless, the primary focus and concern remains on the radiality constraint, which is the most challenging to represent mathematically, as highlighted in \cite{lavorato_imposing_2012}. Other papers \cite{ahmadi_mathematical_2015} propose the use of planar graph representations to ensure radiality while employing a \gls{miqp} formulation. Recent works, such as those published in 2024 \cite{graine_dynamic_2024} and \cite{gallego_mixed-integer_2022}, continue to explore these challenges.

An interesting contribution to the radiality problem was provided by \cite{llorens-iborra_mixed-integer_2012} and \cite{ramos_path-based_2005}. The authors proposed adding a binary signal within a simplified set of equations, along with piecewise linearization, to enable the use of MILP. They also introduced a “path-to-node” incidence concept, which allows for the compact and efficient formulation of both radiality and electrical constraints.

In our work, we have based our implementation on the paper presented by Taylor \cite{taylor_convex_2012}, where the authors propose a set of equations to define radiality similar to \cite{jabr_minimum_2012}, but also introduce a third variable \( y_{ij} \) that defines the state of the line or its associated switch. The problem is then solved by \gls{miqp}, without considering the physical limits, such as minimum bus voltage or maximum line current. In the paper, the authors also propose a second solution, based on a \gls{socp}, where the physical constraints are added to the set of constraints, which has not been implemented in our paper.

The most significant contribution of this proposal \cite{taylor_convex_2012} is the mathematical definition of radiality, where three key variables are introduced:
\begin{itemize}
    \item $z_{ij}$ and $z_{ji}$: To define the current flow direction
    \item $y_{ij}$: To indicate whether the current is active (switch closed)
\end{itemize}

Alongside the radiality constraints, the power flow equations are formulated with different relaxations for each approach:
\begin{itemize}
    \item For \gls{miqp}, the equations consider only resistance (a simplification of the full impedance model).
    \item For \gls{socp}, the equations are extended to incorporate the full impedance model.
\end{itemize}

\subsection{Heuristic techniques}\label{heuristic-techniques}
Heuristic methods leverage domain-specific knowledge of distribution systems to efficiently find feasible configurations, though they typically yield locally optimal solutions. These approaches offer computational advantages over metaheuristics by incorporating electrical system insights directly into their decision-making processes, as will be explained in this section.

The literature classifies heuristic DNR methods into three principal approaches:

\begin{itemize}
    \item Loop Cutting: Systematic opening of switches in meshed networks, starting with all switches closed and prioritizing branches by current magnitude.

    \item Branch Exchange: Iterative switch swapping from an initial configuration. It operates on an existing radial topology and it uses sensitivity metrics for switch selection.

    \item Greedy Graph Methods: Utilize methods like Kruskal's \cite{Kruskal1956} or Prim's \cite{Prim1957} algorithms to find a minimum spanning tree or radial configuration based on electrical parameters such as cable impedance or current or power in a given configuration. The decision-making process in these methods can be specifically tailored by the author or implemented using standard algorithms.

\end{itemize}

The first proposed \gls{dnr} method  was a heuristic loop cutting algorithm by Merlin et al. \cite{merlin_search_1975} in 1975, which selects branches for opening based solely on their minimum current. After selecting a branch, the algorithm ensures no  nodes remain disconnected and repeats the process (see Fig. \ref{fig_merlin}). This method was later refined by \cite{shirmohammadi_reconfiguration_1989}, incorporating AC power flow calculations and network constraints, for particular use during network planning phase.

Two other key studies,  conducted by \cite{civanlar_distribution_1988} in 1988 and \cite{baran_network_1989} in 1989, were based on the Branch Exchange method. These works innovated both in heuristic reconfiguration methods and, more importantly, in power flow estimation techniques. These advancements were crucial at a time when computational capacities were limited, accelerating decision-making processes. In \cite{baran_network_1989}, the author dedicates a significant portion to defining simplified DistFlow equations, as explained in Section \ref{sec:powerflow}. Later, in 1992, \cite{goswami_new_1992} improved the method for selecting the lines to be opened.

The Branch Exchange algorithm has continued to be studied and updated to this day, as evidenced by recent papers such as \cite{salkuti_effective_2021}, where authors use BIBC/BCBV matrices to speed up load flow calculations (similar to Civanlar's approach \cite{civanlar_distribution_1988}), and \cite{pereira_distribution_2023}, published in 2023.

Another area of study involves greedy methods for achieving minimum spanning trees based on graph theory, which aim to select the optimal branches for opening or closing based on localized decision parameters. These methods employ algorithms like Prim's and Kruskal's, as seen in \cite{t_d_loss_2014} and \cite{mosbah_optimum_2017}, where authors use decision parameters ranging from line impedance to line active power.

This paper has selected and implemented five different heuristic methods, spanning the three approaches listed above, as they demonstrate fundamentally different strategies for solving the problem. This contrasts with mathematical methods, where changes primarily involve the description of constraints while the core problem remains the same:
\begin{itemize}
    \item Loop Cutting: The first implemented method in this review is the proposal by Merlin \cite{merlin_search_1975}  (shown in Figure \ref{fig_merlin}). This method starts with all switches closed; the algorithm then calculates the network power flow and lists all branches in ascending order based on their current. It then chooses branches starting from the one with the lowest current and disconnects them sequentially, ensuring that the resulting network maintains radiality. After disconnecting a branch and generating a valid network configuration, the power flow is recalculated, and the process repeats until radiality is reached (\ref{eq:radiality_1}).

    \item Branch Exchange: The selected Branch Exchange methods are based on the algorithms proposed by Baran \cite{baran_network_1989}, represented in Fig.~\ref{fig:baran} and
    the method proposed by Salkuti \cite{salkuti_effective_2021}, represented in Fig.~\ref{fig_saltoki}, and which builds upon earlier proposals by \cite{baran_network_1989} and \cite{civanlar_distribution_1988}.
The main feature of this last algorithm, compared to the other implemented algorithms, is that its optimization objective is to improve the voltage profile rather than directly reducing losses. It starts with a valid radial configuration and iteratively exchanges tie switches. It compares the voltages at the nodes connected by the tie switch, and the algorithm opens the line connected to the node with the higher voltage.
One of the key elements is the use of BIBC/BCBV matrices for efficient power flow calculation in distribution networks.
The authors start with a network configuration compliant with constraints and perform a power flow calculation to determine the voltages at each node. They then analyze each tie switch by comparing the voltages at its connected nodes. Depending on whether the sending or receiving node has the higher voltage, the algorithm decides on the exchange of the tie switch to optimize network performance. The power flow is recalculated after each change, and if the system performance improves, the change is retained; otherwise, the algorithm proceeds to the next tie switch.
\begin{figure}[pos=htbp]
    \centering
    \includegraphics[width=0.65\linewidth]{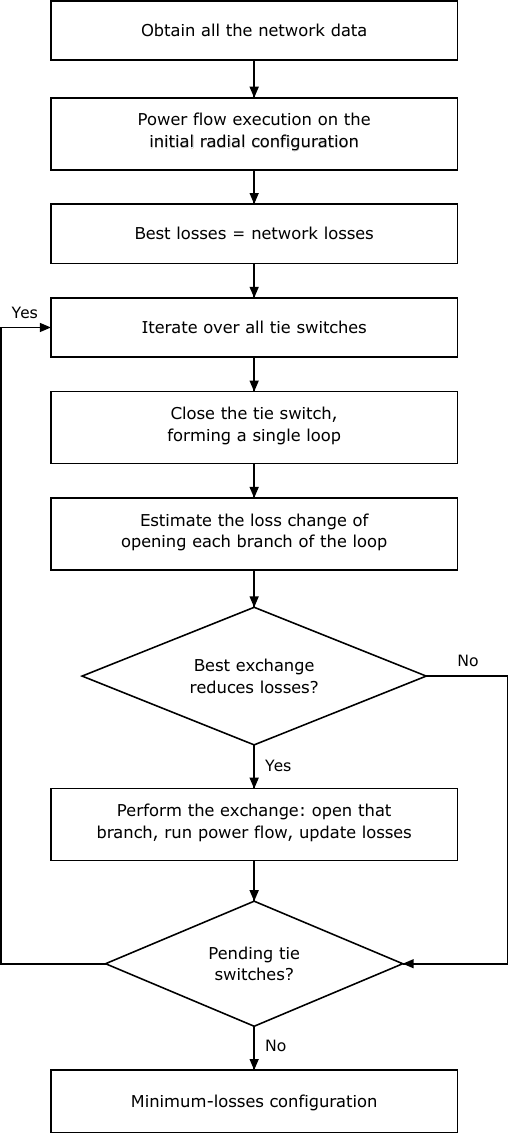}
    \caption{Flow diagram of "Network reconfiguration in distribution systems for loss reduction and load balancing" by Baran and Wu \cite{baran_network_1989}}
    \label{fig:baran}
\end{figure}

    \item Greedy Minimum Spanning Tree: The authors in \cite{montoya_minimal_2012} propose a graph-based method using Kruskal's algorithm to find the minimum spanning tree (MST) for \gls{dnr}. The algorithm begins by calculating the power flow with all switches closed and assigns weights to graph edges based on the inverse of the active power of the branches. Kruskal's MST algorithm is then applied to these weighted edges, which progressively adds edges to the spanning tree based on their weights, ensuring that no cycles are created. This process is illustrated in Fig. \ref{fig_montoya}.
The authors experimented with various weighting policies and found that using the inverse of active power produced comparable results to more complex metaheuristic algorithms such as genetic algorithms and harmony search.

The implementation performed in the present paper is based on \cite{montoya_minimal_2012} but includes the option to use different MST algorithms, such as Prim's or Boruvka's, either the branch current or the active power can be used as the edge weight, having the best results with active power.

\item Brute Force : An alternative heuristic approach is proposed by \cite{morton_efficient_2000}, that  exhaustively search all valid spanning trees in a network using a brute-force strategy. This ensures that the optimal radial solution is identified. Morton's method simplifies power flow calculations by converting the model into a constant-current load-based representation. This modification facilitates a more efficient evaluation of network configurations while retaining the advantages of a spanning tree approach.

\end{itemize}

\begin{figure}[pos=htbp]
    \centering
    \includegraphics[width=1.0\linewidth]{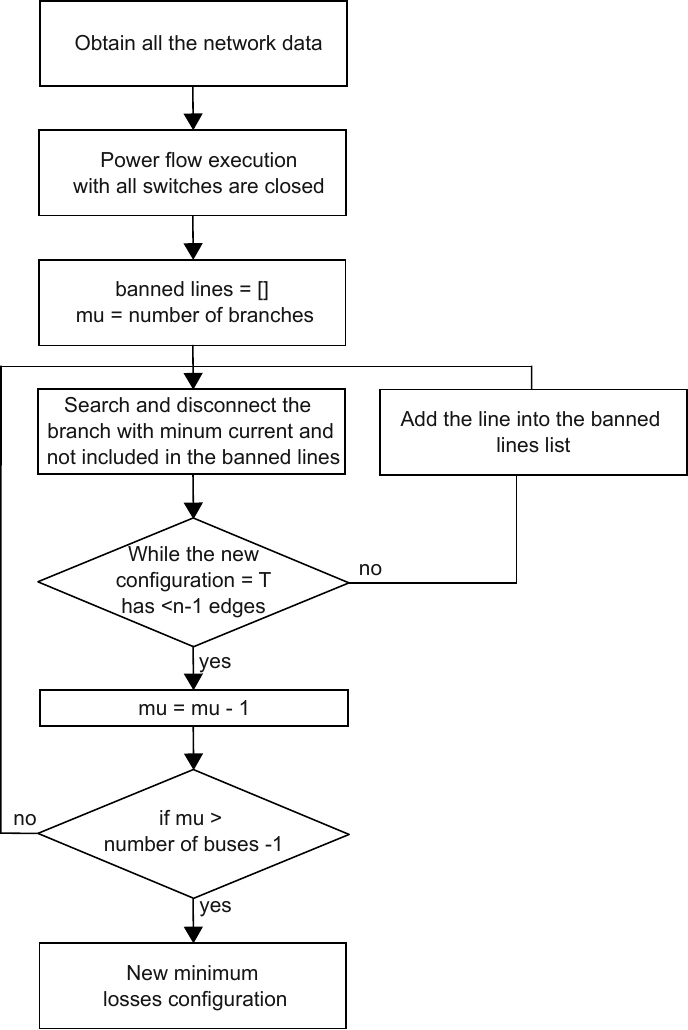}
    \caption{Heuristic algorithm based on "Search for a minimal loss
operating spanning tree configuration in an urban power distribution system" by Merlin et al. \cite{merlin_search_1975}}.
    \label{fig_merlin}
\end{figure}

\begin{figure}[pos=htbp]
    \centering
    \includegraphics[width=1\linewidth]{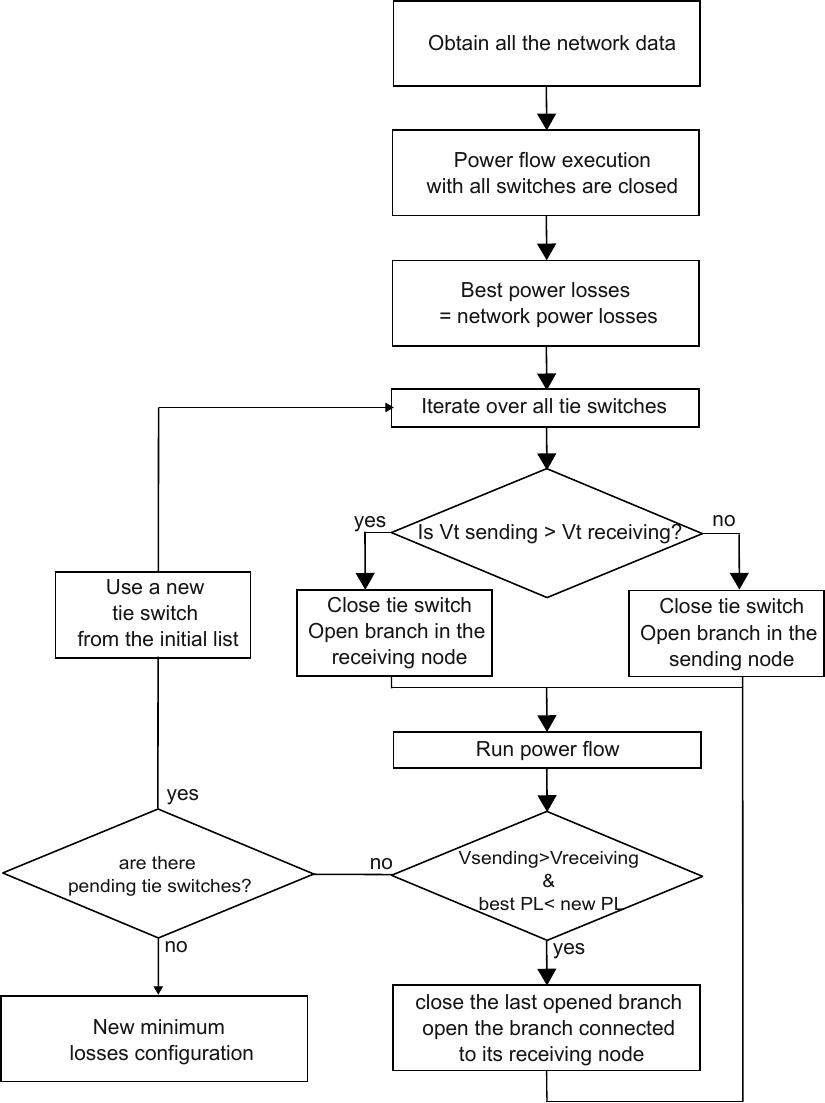}
    \caption{Flow diagram of "An effective network reconfiguration approach of radial distribution system for loss minimization and voltage profile improvement" by Salkuti et al. \cite{salkuti_effective_2021}}
    \label{fig_saltoki}
\end{figure}

\begin{figure}[pos=htbp]
    \centering \includegraphics[width=0.65\linewidth]{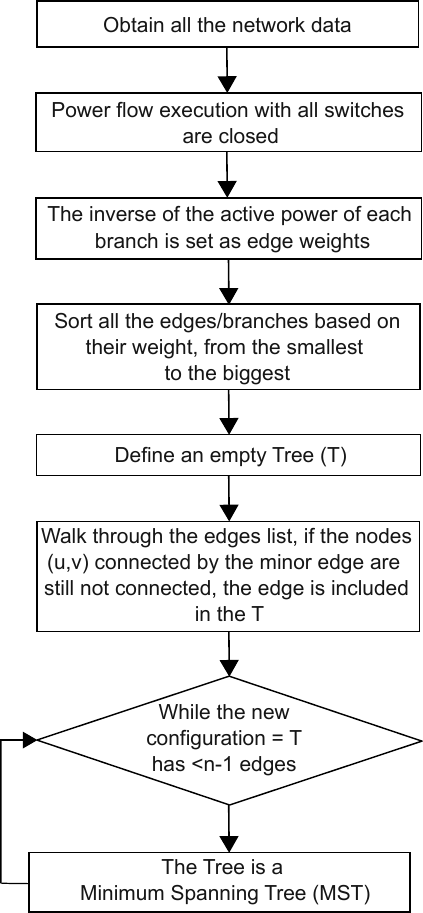}
    \caption{Flow diagram of "A minimal spanning tree algorithm for distribution network configuration" by Montoya et al. \cite{montoya_minimal_2012}}
    \label{fig_montoya}
\end{figure}

\subsection{Metaheuristic Methods}\label{metaheuristic}

Metaheuristic algorithms constitute a significant portion of research in \gls{dnr}, accounting for over 48\% of studies according to \cite{mahdavi_reconfiguration_2021}.

Metaheuristic methods iteratively optimize problems without using derivative information, thereby simplifying computational processes. They often emulate natural processes like genetic evolution or the behaviors of animal collectives (e.g., bird swarms, ant colonies) to find near-optimal solutions.

The pioneering metaheuristic algorithm for DNR was the Genetic Algorithm (GA), which evolves an initial population through crossover, mutation, and selection based on a fitness function to minimize or maximize objectives. GA was first applied to DNR in 1992 \cite{nara_implementation_1992} and gained popularity in the 2000s with refinements such as faster power flow calculations \cite{zhu_refined_1998} and multi-objective approaches \cite{jakus_optimal_2020}.

Population-based algorithms such as Ant Colony Optimization (ACO) and Particle Swarm Optimization (PSO) iteratively improve solutions based on candidate positions and values, converging towards optimal solutions.

Other metaheuristic algorithms applied to DNR include Simulated Annealing, Immune Algorithms, Plant Growth Simulation Algorithm, Honey Bee Mating Optimization, Artificial Bee Colony Algorithm, and Gravitational Search Algorithm, among others. Recent studies have explored newer algorithms such as Harris Hawk Optimization and Jaya Algorithm \cite{helmi_efficient_2022,debbarman_application_2022}.

Despite their effectiveness, metaheuristic algorithms heavily depend on chosen parameters and the initial conditions, which can influence their performance significantly, but also, they are probabilistic approach, so their solution varies from execution to execution.

\begin{figure}[pos=htbp]
    \centering
    \includegraphics[width=0.65\linewidth]{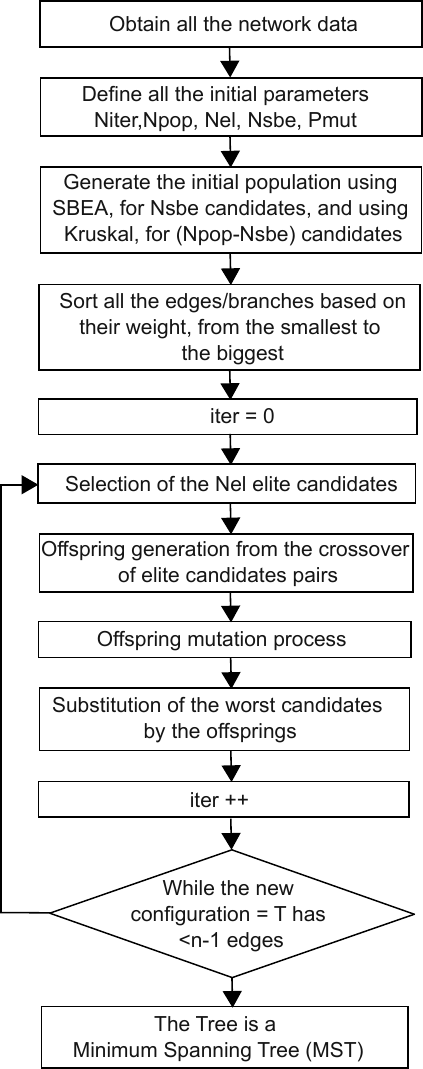}
    \caption{Genetic Algorithm based on the paper "Optimal Reconfiguration of Distribution Networks Using Hybrid Heuristic-Genetic Algorithm" by Jakus et al. \cite{jakus_optimal_2020}}
    \label{fig:fig_jakus}
\end{figure}

\begin{figure}[pos=htbp]
    \centering
    \includegraphics[width=0.75\linewidth]{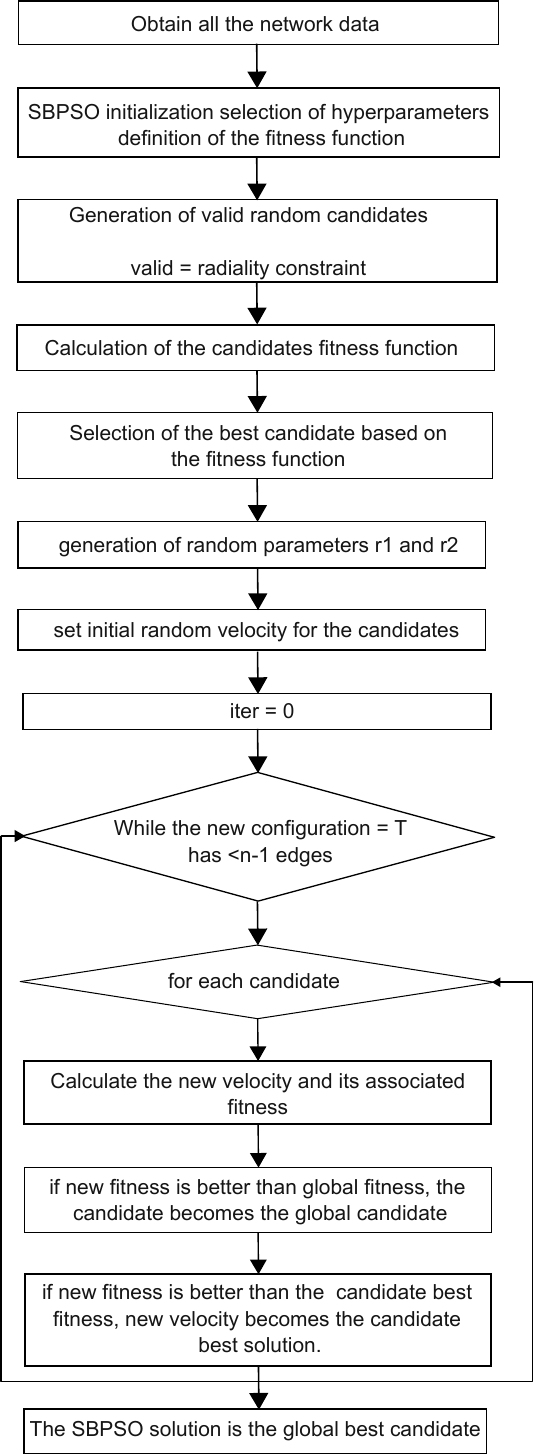}
    \caption{Selective Binary Particle Swarm Optimization (SBPSO) algorithm based on  “Reconfiguration for loss reduction of distribution systems using selective particle swarm optimization” by Khalil et al. \cite{khalil_reconfiguration_nodate}}
    \label{fig_gorpinich}
\end{figure}

For this study, the following two metaheuristic methods were selected:
\begin{itemize}
    \item Genetic Algorithm (GA) \cite{jakus_optimal_2020}: The paper by Jakus \cite{jakus_optimal_2020} introduces a traditional Genetic Algorithm (GA), illustrated in Fig. \ref{fig:fig_jakus}, and described in detail below.

The algorithm begins by generating an initial population of \( N_{pop} \) candidates. Specifically, \( N_{sbe} \) candidates are derived from an initial radial configuration using the Successive Branch Exchange Algorithm (SBEA) based on \cite{baran_network_1989}. The remaining candidates are generated by a greedy minimum spanning tree method employing Kruskal's algorithm, where each branch is assigned a random weight for each candidate to ensure a random exploration of the action space.

Once the initial population is established, each candidate's fitness function is evaluated. The fitness function is defined as the inverse of the power loss plus one to prevent division by zero. Candidates are sorted based on their fitness values from highest to lowest, and the top \( N_{el} \) elite candidates are selected to proceed. The remaining candidates are replaced with new candidates generated through crossover operations between elite candidate pairs, followed by mutations with a probability of (\( P_{mut} \)).

This iterative process continues for \( N_{iter} \) iterations, and the candidate with the highest fitness is selected as the solution to the \gls{dnr} problem.
    \item Selective Binary Particle Swarm Optimization (SBPSO) \cite{khalil_reconfiguration_nodate}:
The Particle Swarm Optimization (PSO) algorithm was first introduced by \cite{kennedy_particle_1995} and has been applied to the \gls{dnr} problem since the 2000s. Initial works such as \cite{jin_distribution_2004} utilized the Binary Particle Swarm Optimization (BPSO) formulation, which was later enhanced by \cite{khalil_reconfiguration_nodate} by introducing a technique for efficiently selecting the line to disconnect. Comparative studies such as \cite{7019093} provide valuable insights into different PSO implementations.

The algorithm proposed by Khalil \cite{khalil_reconfiguration_nodate}, depicted in Fig. \ref{fig_gorpinich}, starts by gathering all network information, defining configuration parameters, and generating a random list of valid candidates. Each candidate adheres to radiality constraints (i.e., no loops and all nodes connected), and is represented by a list of open switches. The algorithm defines the search dimension as the number of loops to be opened, and the search space is defined by the available switches at each loop. How repeated switches are assigned to one or another loop, in order to avoid non valid solutions, is a key limiting point in order to achieve optimal configurations for this method.

Once the initial candidates are defined, the fitness function is evaluated through power flow calculations. The best global candidate is identified as the one with lowest losses.

Subsequently, the algorithm computes the velocity of each candidate using the equations originally proposed by \cite{kennedy_particle_1995}:

\begin{align}
v_{id}^{k+1} = wv_{id}^k + c_1 r_1 (pb_{id}^k - x_{id}^k) + c_2 r_2 (gb_d^k - x_{id}^k) \label{eq:sbpso1} \\
x_{id}^{k+1} = x_{id}^k + v_{id}^{k+1} \label{eq:sbpso2}\\
\text{sigmoid}(v_{id}^{k+1}) = \frac{1}{1 + \exp(-v_{id}^{k+1})} \label{eq:sbpso3}
\end{align}

where \( id = 1, 2, \ldots, n \) represents the population or set of swarm particles, \(n\) being the number of particles and \(d\) the dimension index; \(x_{id}^k\) is the current position of particle \(id\) at iteration \(k\), \(v_{id}^k\) its velocity, \(pb_{id}^k\) the best position found so far by that particle, and \(gb_d^k\) the best position found by the whole swarm in dimension \(d\); and \( w, c_1, c_2, r_1, r_2 \) are inertia, acceleration and randomness constants, respectively, and \(k\) the iteration up to \(N_{iter}\) iterations.

Authors in \cite{khalil_reconfiguration_nodate} proposed a modification to search within the action space, composed of switches available at each loop, by adjusting the sigmoid equation of BPSO (eq. \ref{eq:sbpso3}), resulting in Selective Binary Particle Swarm Optimization (SBPSO):

\begin{align}
\text{sigmoid}(v_{id}^{k+1}) = d_n \frac{1}{1 + \exp(-v_{id}^{k+1})} \label{eq:sbpso4}
\end{align}

where \( d_n \) represents the search space in the loop, ensuring that the selection of the new line to be opened is conducted appropriately.

\begin{align}
x_{id}^{k+1} =
\begin{cases}
s_{d1} & \text{if } \text{sigmoid}(v_{id}^{k+1}) < 1 \\
s_{d2} & \text{if } \text{sigmoid}(v_{id}^{k+1}) < 2 \\
s_{d3} & \text{if } \text{sigmoid}(v_{id}^{k+1}) < 3  \label{eq:spbso5} \\
\vdots \\
s_{d_n} & \text{if } \text{sigmoid}(v_{id}^{k+1}) < d_n \\
\end{cases}
\end{align}

where $s_{d_n}$ are the selected values in dimension $d$. After selecting a new line to open, the fitness function is recalculated. If any candidate improves upon the global candidate's fitness value, that candidate becomes the new global solution.

This iterative process continues for \(N_{iter}\) iterations, with the final global candidate determined as the algorithm's solution.

\end{itemize}

% =============================================================================

\begin{figure}[pos=htbp]
  \centering
  \includegraphics[width=0.95\linewidth]{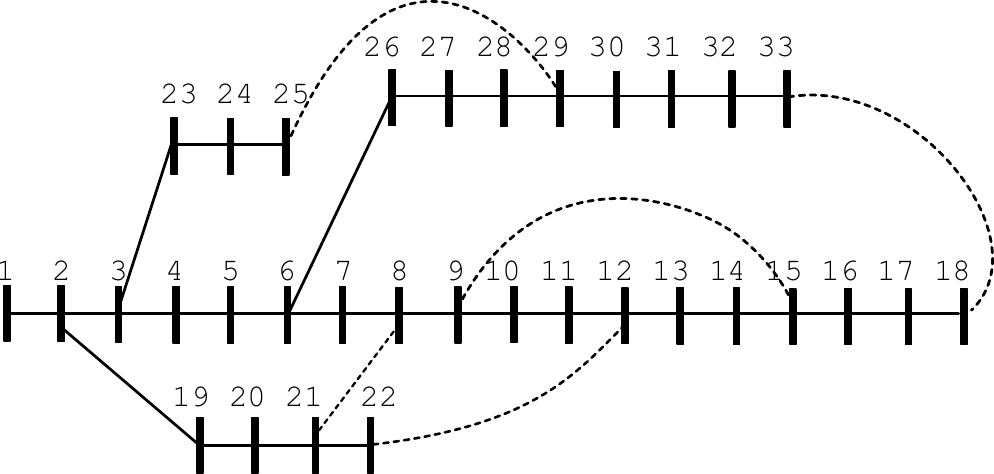}
  \caption{33-bus distribution system \cite{baran_network_1989}. Dashed lines are the default tie lines.}
  \label{fig_33buses}
\end{figure}

\begin{figure}[pos=htbp]
    \centering
    \includegraphics[width=1\linewidth]{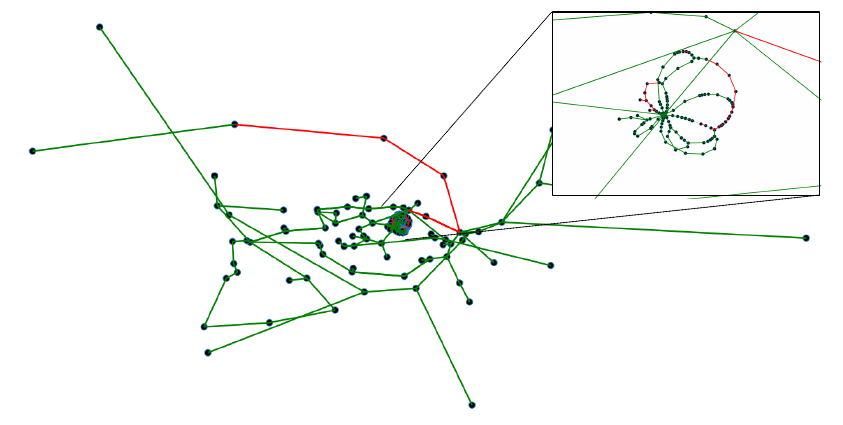}
    \caption{Simbench 1-HVMV-urban-2.203-0-no\_sw network.}
    \label{fig:simbench_network}
\end{figure}

\begin{figure*}[width=\textwidth,pos=tbp]
    \centering
    \includegraphics[width=1\linewidth]{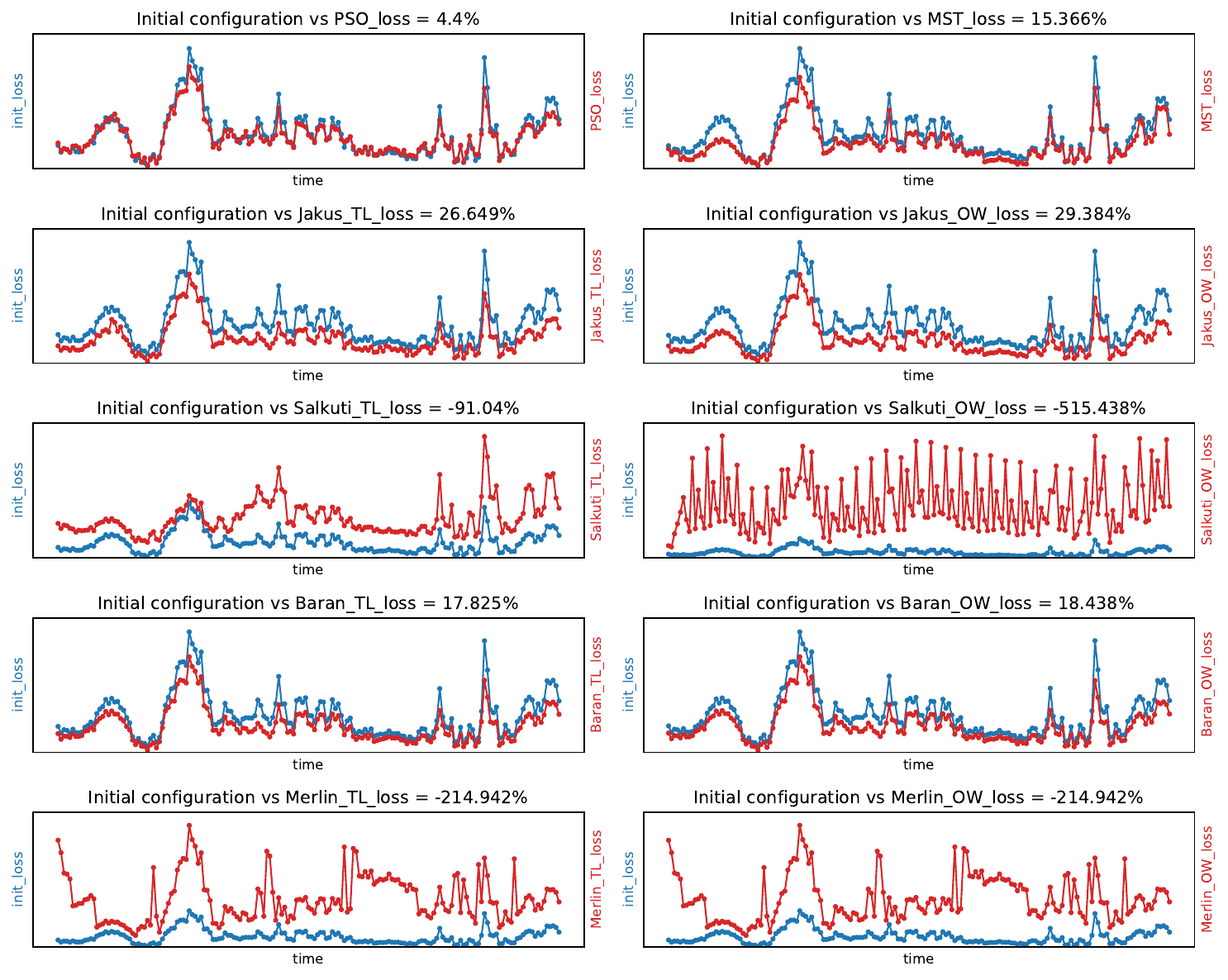}
    \caption{Method performance over the Simbench urban network time series during one week (Initial configuration in blue and optimized configuration in red).}
    \label{fig:Simbench_timeseries}
\end{figure*}

\section{Results and Discussion}
\label{sec:results}

\subsection{Test Systems}
\label{subsec:testsystems}
All algorithms were evaluated on the five networks of Table~\ref{tab:test_systems}.
The first four---the 16-, 33-, 69- and 118-bus systems---are static benchmarks solved at nominal load. The fifth, the 196-bus Simbench urban network \emph{1-HVMV-urban-2.203-0-no\_sw}, carries one full year of 15-minute load and
generation profiles and is used both as the largest static case and for the dynamic study of Section~\ref{subsec:dynamic}. The 33-bus system is shown in Fig.~\ref{fig_33buses} and the Simbench network in Fig.~\ref{fig:simbench_network}.

\begin{table}[pos=htbp]
    \caption{Test networks}
    \label{tab:test_systems}
    \centering
    \footnotesize
    \setlength{\tabcolsep}{3.5pt}
    \begin{tabular}{|l|c|c|c|c|c|}\hline
    \textbf{System} & \textbf{Buses} & \textbf{Lines} & \textbf{Loads} & \textbf{Tie sw.} & \textbf{Source} \\ \hline
    16-bus  & 16  & 20  & 14  & 7 & \cite{civanlar_distribution_1988} (1988) \\ \hline
    33-bus  & 33  & 37  & 33  & 5       & \cite{baran_network_1989} (1989) \\ \hline
    69-bus  & 69  & 74  & 48  & 5       & \cite{Baran1989} (1989) \\ \hline
    118zh-bus & 118 & 132 & 117 & 13 & \cite{ZHANG2007685} (2007) \\ \hline
    Simbench urban & 196 & 215 & 194 & 17 & \cite{Meinecke_Sarajlic_Drauz-Mauel_Klettke_Lauven_Rehtanz_Moser_Braun_2020} (2020) \\ \hline
    \end{tabular}
\end{table}

\subsection{Experimental Setup and Reproducibility}
\label{subsec:setup}
Every algorithm was re-implemented in Python and driven by the same power-flow
engine, which can be chosen by the user between VeraGrid~\cite{VeraGrid_2025} and PandaPower~\cite{pandapower_2018}, so that differences in reported losses
reflect differences in search strategy rather than in the underlying power-flow
approximation. Networks are imported from MATPOWER \texttt{.m} files~\cite{matpower_2011}
or from the Simbench dataset~\cite{Meinecke_Sarajlic_Drauz-Mauel_Klettke_Lauven_Rehtanz_Moser_Braun_2020}. The mathematical formulation of
Section~\ref{sec:mathematical-formulation-and-constraints} is built with Pyomo~\cite{hart2011pyomo,bynum2021pyomo}
and solved with SCIP~\cite{scip_2009}.

\begin{table}[pos=htbp]
    \caption{Algorithm hyperparameters}
    \label{tab:hyperparams}
    \centering
    \footnotesize
    \begin{tabular}{|l|l|c|}\hline
    \textbf{Algorithm} & \textbf{Parameter} & \textbf{Value} \\ \hline
    \multirow{5}{*}{GA \cite{jakus_optimal_2020}}
        & Population size $N_{pop}$      & 16 \\ \cline{2-3}
        & Elite candidates $N_{el}$      & 2 \\ \cline{2-3}
        & Mutation probability $P_{mut}$ & 0.4 \\ \cline{2-3}
        & Generations $N_{iter}$         & 20 \\ \hline
    \multirow{4}{*}{SBPSO \cite{khalil_reconfiguration_nodate}}
        & Swarm size                     & 10 \\ \cline{2-3}
        & Inertia $w$                    & 0.9 - 1.2 \\ \cline{2-3}
        & Acceleration $c_1, c_2$        & 2 \\ \cline{2-3}
        & Iterations $N_{iter}$          & 20 \\ \hline
    \end{tabular}
\end{table}

Table~\ref{tab:hyperparams} lists the hyperparameters of the two stochastic
methods, SBPSO \cite{khalil_reconfiguration_nodate} and GA \cite{jakus_optimal_2020}. Their stochastic nature makes every execution different, Table~\ref{tab:stocastic_variability} shows the average result and the standard variation for the losses and the execution time for 10 runs of each of the methods.

Because both are stochastic, every result reported for them is the mean
over $N_{\mathrm{runs}}$ independent runs with fixed seeds $0,\dots,N_{\mathrm{runs}}-1$, and is reported with its sample standard deviation (Table~\ref{tab:stocastic_variability}). Deterministic methods are run once.

\begin{table}[pos=htbp]
    \caption{Variability on the metaheuristic methods results for the 33-bus case. }
    \label{tab:stocastic_variability}
    \centering
    \footnotesize
    % Fixed-width wrapping columns. With |c|c|c| the single-line headers made
    % this 272pt wide in a 237pt column -- it ran 41pt past the right text
    % margin. It was also the only table left at the class's default \small.
    \setlength{\tabcolsep}{2pt}
    \begin{tabular}{|>{\centering\arraybackslash}p{0.16\linewidth}|>{\centering\arraybackslash}p{0.34\linewidth}|>{\centering\arraybackslash}p{0.40\linewidth}|}
        \hline
        Method & Losses (mean+/-std)(kW)  & Exec.time.  (mean+/-std)(sec) \\ \hline
        Khalil & 168.790 +/- 2.755  & 2806.203 +/- 778.61 \\ \hline
        Jakus  & 140.133 +/- 0.306 & 2982.907 +/- 354.89 \\ \hline
    \end{tabular}
\end{table}

\begin{table}[pos=htbp]
    \caption{Computation cost on the 33-bus system at nominal load}
    \label{tab:33buses_table}
    \centering
    \footnotesize
    \setlength{\tabcolsep}{3pt}
    \begin{tabular}{|c|c|c|c|}\hline
    \textbf{Algorithm} & \textbf{Family} & \textbf{NPF} & \textbf{Time (sec.)}   \\ \hline
    Merlin  & Heuristic     & 6          & 0.121     \\ \hline
    Baran   & Heuristic     & 447          & 6.60      \\ \hline
    Montoya & Heuristic     & 1           & 0.034      \\ \hline
    Morton  & Heuristic     & $>10{,}000$ & hours   \\ \hline
    Salkuti & Heuristic     & 12          & 0.32       \\ \hline
    Khalil  & Metaheuristic & 240         & 4.82   \\ \hline
    Jakus   & Metaheuristic & 543         & 9.02   \\ \hline
    Taylor  & Mathematical  & 0           & $>$120   \\ \hline
    \end{tabular}
\end{table}

Execution time depends on the machine, the language, and the maturity of the implementation, and therefore ages  badly. We report instead the \gls{npf}, Table~\ref{tab:33buses_table}, required to reach the returned solution. Solving the power flow dominates the cost of every method considered here, so NPF captures the algorithmic work while remaining invariant to hardware. NPF is only comparable across methods when the per-evaluation cost is equal; since all methods in this study share one power-flow engine, it is. The one exception is the mathematical method: the MIQP solver evaluates the power balance internally as part of the branch-and-bound tree rather than through explicit power-flow calls, so NPF is not meaningful for it and only time is reported.

\subsection{Loss Minimization on Static Benchmarks}
\label{subsec:staticloss}
Table~\ref{tab:33_losses} reports the losses obtained on the 33-bus system, and Fig.~\ref{fig:loss_vmin} shows losses and minimum voltage across all four static networks.

\begin{table}[pos=htbp]
    \caption{Losses on the 33-bus system}
    \label{tab:33_losses}
    \centering
    \footnotesize
    \setlength{\tabcolsep}{3pt}
    \begin{tabular}{|l|l|l|c|}\hline
    \textbf{Author} & \textbf{Family} & \textbf{Technique} & \textbf{Losses (kW)} \\ \hline
    Merlin \cite{merlin_search_1975}   & Heuristic     & Loop cutting     & 140.27 \\ \hline
    Baran \cite{baran_network_1989}    & Heuristic     & Branch exchange  & 139.55  \\ \hline
    Montoya \cite{montoya_minimal_2012}& Heuristic     & Greedy MST       & 140.27  \\ \hline
    Morton \cite{morton_efficient_2000}& Heuristic     & Brute Force       & 139.55  \\ \hline
    Salkuti \cite{salkuti_effective_2021} & Heuristic  & Branch exchange  & 139.55
    \\ \hline
    Khalil \cite{khalil_reconfiguration_nodate} & Metaheuristic & SBPSO   & 172.03 \\ \hline
    Jakus \cite{jakus_optimal_2020}    & Metaheuristic & GA + SBEA        & 139.55 \\ \hline
    Taylor \cite{taylor_convex_2012}    & Mathematical & MIQP        & 139.55 \\ \hline    \end{tabular}
\end{table}

On the 33-bus system, seven of the eight methods converge to essentially the same minimum, $139.55$--$140.27$~kW: the two branch-exchange methods (Baran, Salkuti), the genetic algorithm (Jakus) and the mathematical program (Taylor) all reach $139.55$~kW, while the loop-cutting (Merlin) and greedy-MST (Montoya) heuristics
settle one tie-line away at $140.27$~kW. The SBPSO (Khalil) is the only method that ends noticeably higher, at $172.03$~kW, reflecting the tendency of population-based search to stall on a sub-optimal spanning tree when the number of loops is large, but also due to its strong dependence on the initial conditions. The minimum losses of $139.55~kW$ are validated by using the brute force algorithm proposed by \cite{morton_efficient_2000}.

Across the four static networks, most of methods reduce losses relative to the initial tie-line configuration, and loss reduction is generally accompanied by an improvement in the minimum bus voltage---the two objectives are aligned on these radial feeders. The exception is Salkuti, whose objective is the voltage profile rather than losses (Section~\ref{heuristic-techniques}):
it improves the minimum voltage competitively but leaves losses close to the initial configuration.

\subsection{Computational Cost}
\label{subsec:cost}

\begin{figure}[pos=htbp]
    \centering
    \includegraphics[width=0.85\linewidth]{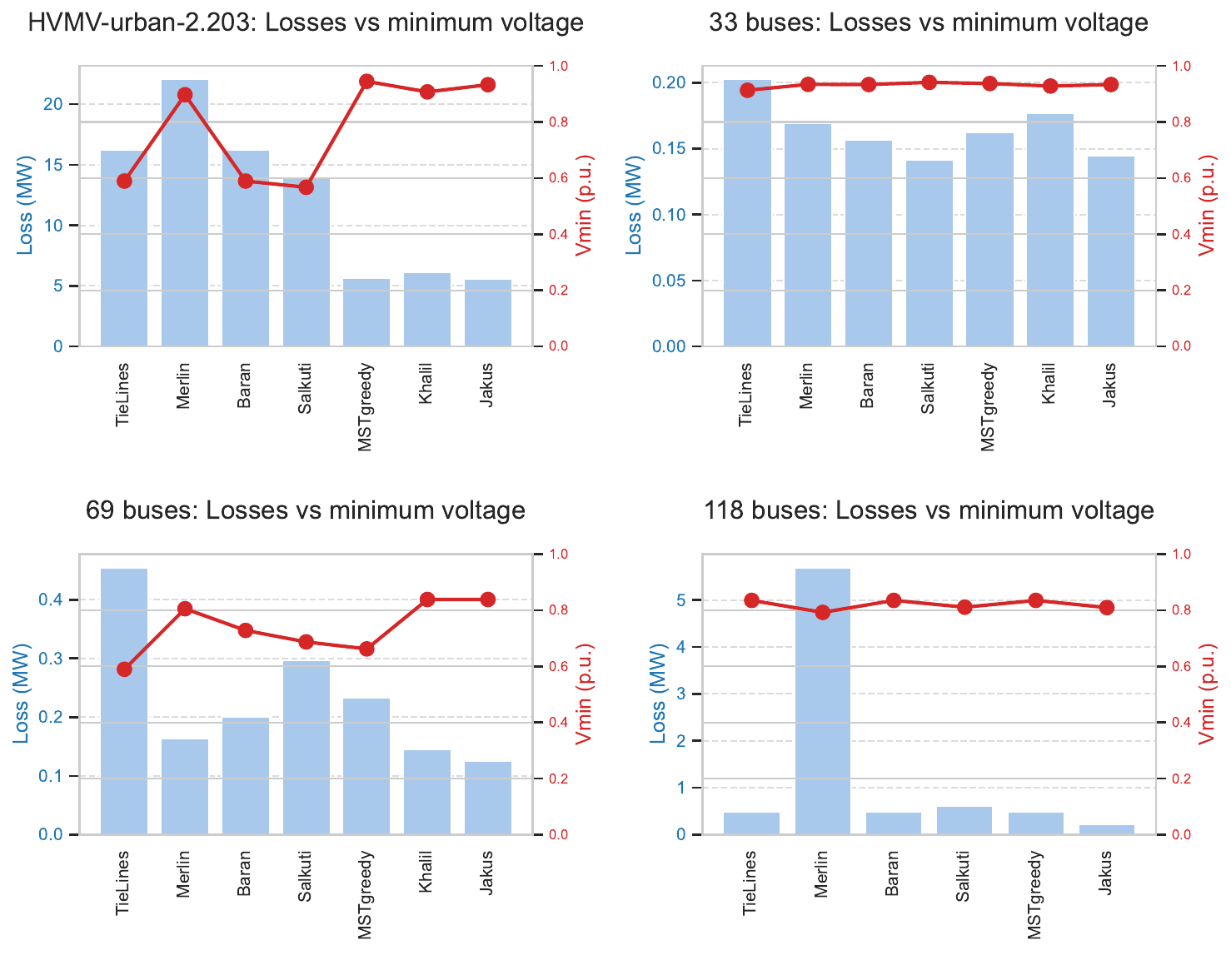}
    \caption{Losses, minimum bus voltage and execution time for each method on the four static test systems.}
    \label{fig:loss_vmin}
\end{figure}

Fig.~\ref{fig:loss_vmin} pairs losses with execution time for the four static networks, two groups emerge, the three of the heuristics (Merlin, Salkuti and Montoya) reach their solution almost instantly, since each performs only a small, bounded number of power flows to obtain the final solution; the greedy MST is the fastest of all, as it requires a single power flow to compute its edge weights plus one to report the result, while Baran branch exchange method can be very time consuming, not achieving its solutions in some of the cases. The metaheuristics (Khalil, Jakus) sit an order of magnitude higher in time because they evaluate a whole population per generation, and the exact MIQP (Taylor) is the slowest by a wide margin, on the larger cases it exhausts its time budget of 120~s without closing the optimality gap, even though it typically finds the best feasible solution. The practical reading of Fig.~\ref{fig:loss_vmin} is that, for these networks, the heuristics dominate the accuracy--cost trade-off: they reach the same loss as the exact method at a fraction of the computational effort.

\subsection{Voltage Profile}
\label{subsec:voltage}

\begin{figure}[pos=htbp]
  \centering
  \includegraphics[width=0.95\linewidth]{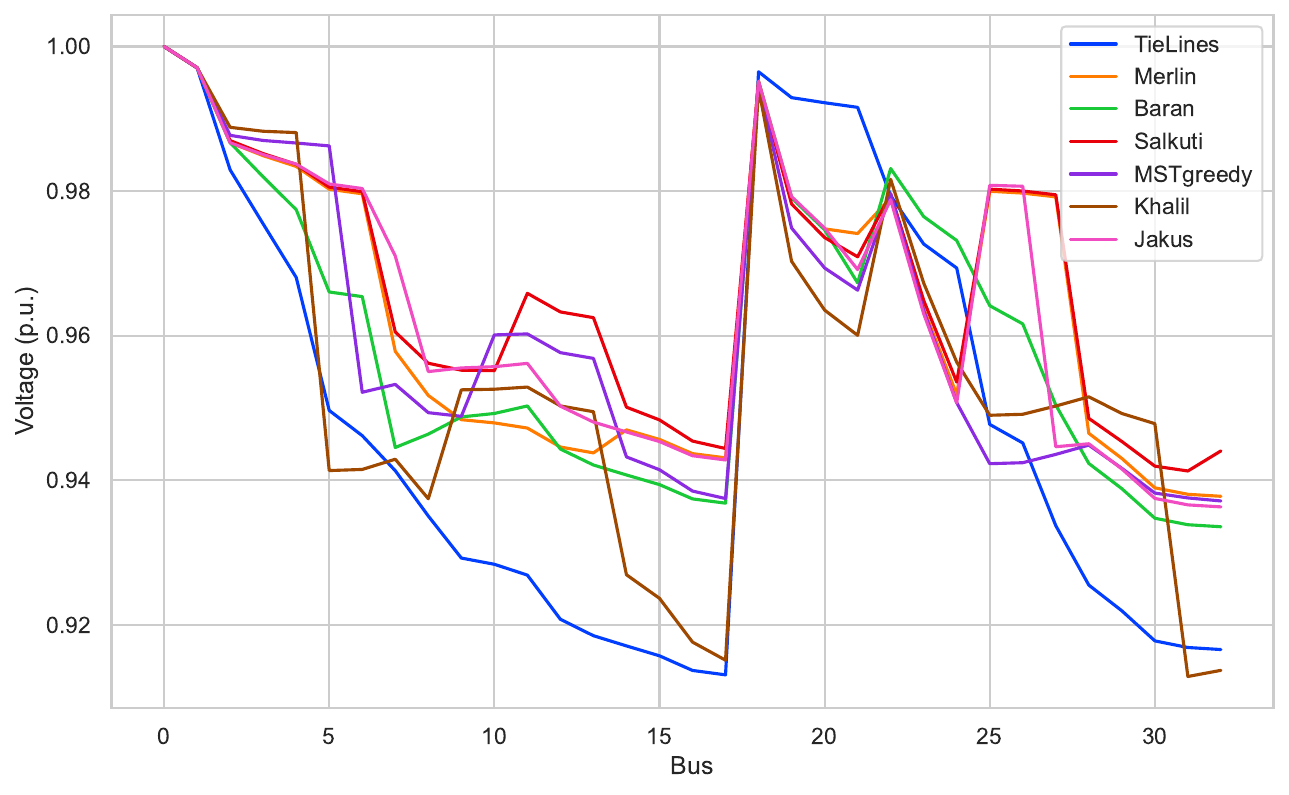}
  \caption{Voltage profile of the 33-bus system before and after reconfiguration.}
  \label{fig:33buses_voltage_GA}
\end{figure}

Fig.~\ref{fig:33buses_voltage_GA} shows the bus voltage profile of the 33-bus system for every method. The initial tie-line configuration sags to a minimum of approximately $0.913$~p.u. at the feeder end, whereas all reconfiguration methods
lift the worst bus to roughly $0.94$--$0.95$~p.u. Most methods produce nearly coincident profiles because they converge to the same or neighbouring spanning trees; Montoya and Khalil deviate slightly, trading a marginally lower voltage on
one lateral for gains elsewhere. This confirms that, on radial distribution feeders, minimizing losses and flattening the voltage profile are largely complementary objectives.

\subsection{Dynamic Study}
\label{subsec:dynamic}
\begin{figure}[pos=htbp]
    \centering
    \includegraphics[width=1\linewidth]{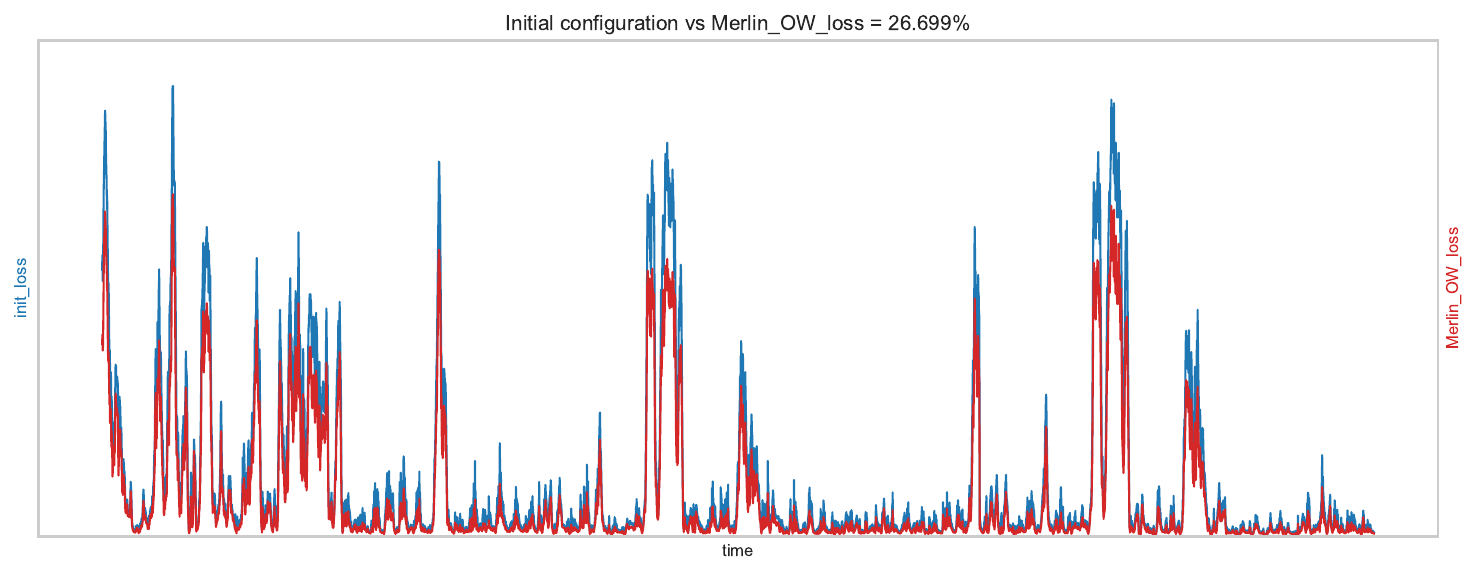}
    \caption{One year hourly timeseries for Simbench network optimized with Merlin method}
    \label{fig:one_year_timeseries}
\end{figure}

We run a dynamic reconfiguration policy over the Simbench urban network, sampled hourly over a representative week. At each time step the network is re-optimized, and we
compare two policies: \emph{TL}, in which every method restarts from the fixed initial tie-line set, and \emph{OW}, in which each method warm-starts from its own solution at the previous time step. Fig.~\ref{fig:Simbench_timeseries} shows, for each method, the instantaneous losses of the reconfigured network (red) against those of the static initial configuration (blue) over the week, together with the mean loss reduction.

Three observations follow. First, the loss-minimizing methods recover a large and
stable fraction of the losses throughout the week: Baran ($25.6\%$), the genetic algorithm ($24.6\%$ TL, $25.5\%$ OW), Merlin ($22.9\%$), the greedy MST ($22.7\%$) and Taylor ($22.5\%$) all track the load closely, confirming that a
single optimized topology is far from optimal once load and generation vary. Second, Salkuti recovers almost nothing ($0.4$--$2.0\%$): because its objective is the voltage profile, it barely moves the loss, which is the expected behaviour and a useful negative control. Third, warm-starting helps exactly the methods that are
stochastic: the genetic algorithm improves from $24.6\%$ (TL) to $25.5\%$ (OW), whereas the deterministic methods (Baran) return identical results under TL and OW, since their outcome does not depend on the starting point.

Over the full year (Fig. \ref{fig:one_year_timeseries}) we have applied the Merlin's loop cutting heuristic method, as it represents a good trade off between savings and execution time, this policy recovers more than 26\% of the total losses, 3.742~MWh at the cost of 55845 switching operations, a figure that motivates the discussion of Section~\ref{subsec:cblife}.

\subsection{Circuit-breaker lifespan under dynamic DNR}
\label{subsec:cblife}

\begin{table}[pos=htbp]
    \caption{Dynamic DNR on the Simbench urban network}
    \label{tab:timeseries_switching}
    \centering
    \footnotesize
    \setlength{\tabcolsep}{2pt}
    \begin{tabular}{|>{\raggedright\arraybackslash}p{0.28\linewidth}|>{\centering\arraybackslash}p{0.25\linewidth}|>{\centering\arraybackslash}p{0.19\linewidth}|>{\centering\arraybackslash}p{0.20\linewidth}|}\hline
    \textbf{Method} & \textbf{Loss reduction (\%)} & \textbf{Time (sec)/step} & \textbf{Switch ops/day} \\ \hline
    Merlin          & 22.94 & 2.26   & 153 \\ \hline
    Baran           & 25.61 & 155.86  & 151 \\ \hline
    Montoya (MST)   & 22.67 & 0.51   & 191 \\ \hline
    Salkuti         & 1.99  & 12.51    &  79 \\ \hline
    Khalil (SBPSO)  & 19.18 & 20.36    & 557 \\ \hline
    Jakus (GA)      & 25.52 & 29.09   & 114 \\ \hline
    Taylor          & 22.50 & 120.00  & 446 \\ \hline
    \end{tabular}
\end{table}

The main obstacle to deploying a dynamic \gls{dnr} strategy at DSOs is the
accelerated ageing of the switching devices: a circuit breaker tolerates a
limited number of operations before it must be replaced, rated at
$N_{\mathrm{life}} = 10\,000$ operations for the class~M2 mechanical endurance of
IEC~62271-100~\cite{iec_62271_100}. Every reconfiguration step therefore consumes
a fraction of that budget, and the resulting maintenance cost must be weighed
against the energy savings that motivate the strategy in the first place.

\begin{figure}[pos=htbp]
    \centering
    {\includegraphics[width=1\linewidth]{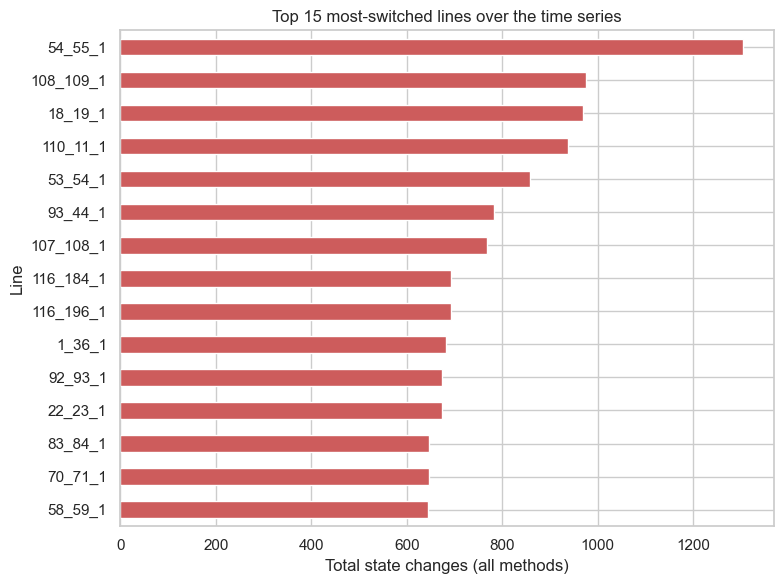}}%
    \caption{Most switched circuit breakers for one year time series, for Simbench case, using Merlin method}
    \label{fig:most_switched_Cb}
\end{figure}

Let $C_{cb}$ be the total cost of repairing/replacing a circuit breaker, including both the device/parts
and labour cost. Amortizing this
cost over the breaker's rated endurance gives the cost attributable to a single
switching action ($C_{\mathrm{sw}}$),
\begin{equation}
    C_{\mathrm{sw}} = \frac{C_{cb}}{N_{\mathrm{life}}}.
    \label{eq:cost_switch}
\end{equation}
If a dynamic \gls{dnr} policy performs $n_{\mathrm{op}}$ switching operations over
a given period, its incremental switching cost is
\begin{equation}
    C_{\mathrm{dDNR}} = n_{\mathrm{op}}\,C_{\mathrm{sw}}
                     = n_{\mathrm{op}}\,\frac{C_{cb}}{N_{\mathrm{life}}}.
    \label{eq:cost_dnr}
\end{equation}
Denoting by $S_{E}$ the economic cost of the energy losses savings over the same
period, the policy is economically justified when the switching cost stays below
the savings it enables,
\begin{equation}
    n_{\mathrm{op}}\,\frac{C_{cb}}{N_{\mathrm{life}}} < S_{E}.
    \label{eq:breakeven}
\end{equation}
Equation~\eqref{eq:breakeven} makes the trade-off explicit: a method is worthwhile
not merely when it reduces losses, but when it does so with few enough switching
operations that the saved energy outweighs the accelerated replacement of the
breakers. This favours methods that reach a low-loss configuration with a small,
stable set of switch changes over those that churn many switches for a marginal
loss reduction (Section~\ref{subsec:dynamic}). Fig.~\ref{fig:most_switched_Cb} shows the 15 most switched lines for the one year timeseries shown in Fig.~\ref{fig:one_year_timeseries}, while Table~\ref{tab:timeseries_switching} compares the loss reduction achieved by each method, with the calculation time and the number of circuit breaker operations per day.

% =====================================================================
\section{Software}
\label{sec:software}
All methods, test cases and result scripts are released as an open-source Python
library~\cite{dnr_github_2025} so that every figure and table in this paper can be regenerated from a single command.

\subsection{Architecture}
The library is organized around a solver-agnostic facade,
\texttt{DistributionNetworkReconfiguration}, exposing a single entry point,
\texttt{Solve(method=\dots)}, Listing~\ref{lst:DNR_library}, that dispatches to the eight algorithms. Each algorithm is implemented as a self-contained class following a common calling convention: it receives a grid and a set of tie lines along with hyperparameters, returns the list of branches to open, and exposes an \texttt{NPF} counter so that the cost metric of
Section~\ref{subsec:setup} is measured uniformly. Two parallel back-ends are provided and kept behaviorally identical: a VeraGrid-native and a Pandapower-native one, so that results can be
cross-checked between two independent power-flow engines.

\subsection{Methods}
The eight solvers are: Merlin (loop cutting)~\cite{merlin_search_1975}, Baran and Salkuti
(branch exchange)~\cite{baran_network_1989,salkuti_effective_2021}, Montoya (greedy minimum
spanning tree)~\cite{montoya_minimal_2012}, Morton (exhaustive spanning-tree
search)~\cite{morton_efficient_2000}, Khalil (selective binary PSO)~\cite{khalil_reconfiguration_nodate}, Jakus
(genetic algorithm)~\cite{jakus_optimal_2020}, and Taylor (MIQP)~\cite{taylor_convex_2012}. The first seven use the shared VeraGrid or PandaPower Newton-Raphson
power flow and radiality check; the Taylor method uses Pyomo library, \cite{hart2011pyomo,bynum2021pyomo} for formulating constraint and objective functions, and the SCIP solver \cite{scip_2009} to obtain the final solution.

\begin{figure}[pos=htbp]
    \begin{lstlisting}[caption={Example of SBPSO algorithm call for the 69-bus case, using VeraGrid}, label={lst:DNR_library}, basicstyle=\ttfamily\footnotesize]
from VeraGridEngine.IO.file_handler import FileOpen

gridGC = FileOpen("case69.m").open()
dnr = DistributionNetworkReconfiguration(gridGC)
disabled_lines = dnr.Solve(method="Khalil", NumCandidates=10)

_, losses = GC_utils.GC_PowerFlow(gridGC, config=DisabledLinesID)
radiality = GC_utils.CheckRadialConnectedNetwork(gridGC)

    \end{lstlisting}
\end{figure}

\subsection{Reproducibility}
The repository ships the five test networks, a regression-test suite (\texttt{pytest})
covering every solver, the facade and the shared utilities, and a set of example
scripts and notebooks that generate the results of Section~\ref{sec:results}:
a static summary across all cases, the year-long Simbench time-series study, the
voltage-profile comparison, and the loss--time trade-off plots. Fixed random
seeds make the stochastic results reproducible.

% =====================================================================
\section{Conclusion}
\label{sec:conclusion}
This paper re-implemented eight representative (summarized in Table \ref{tab:qualitative_summary}) DNR algorithms, spanning three of
the four established solution paradigms, in a single open-source framework and
benchmarked them on five networks under identical conditions, using the number of
power-flow evaluations as a hardware-independent cost metric.

\begin{table}[pos=htbp]
    \caption{Qualitative comparison of the reviewed methods}
    \label{tab:qualitative_summary}
    \centering
    \footnotesize
    \begin{tabular}{|>{\centering\arraybackslash}m{0.25\linewidth}|>{\raggedright\arraybackslash}m{0.65\linewidth}|} \hline
         \textbf{Morton} (brute force) & Globally optimal; intractable beyond $\approx$33 buses. \\ \hline
         \textbf{Merlin} & Single-pass loop cutting. Historically first; not competitive. \\ \hline
         \textbf{Baran} & Branch exchange. Deterministic, strongly topology-dependent. \\ \hline
         \textbf{Salkuti} & Branch exchange driven by nodal voltage. Best minimum voltage among heuristics. \\ \hline
         \textbf{Montoya} (Greedy MST) & One power-flow evaluation; fastest by two orders of magnitude. \\ \hline
         \textbf{Khalil} (SBPSO) & Fast per iteration; loop-based encoding makes it initialization-dependent and fragile at scale. \\ \hline
         \textbf{Jakus} (GA) & Best solution quality on every case; two to three orders of magnitude more power-flow evaluations. \\ \hline
         \textbf{Taylor} (MIQP) & Globally optimal within the simplified-DistFlow (QP) model; requires a mixed-integer solver.  \\ \hline
    \end{tabular}
\end{table}

Three conclusions stand out. First, on the static benchmarks the simple
branch-exchange and loop-cutting heuristics reach the same loss as the exact
mixed-integer program at a small fraction of its computational cost; the exact
method is valuable chiefly as a certificate that the heuristics are at or near the
global optimum, not as an operational tool for these network sizes. Second, the
choice of objective is decisive and must be stated explicitly: a voltage-driven
method such as Salkuti improves the voltage profile but should not be judged on
losses, on which it is nearly inert. Third, the year-long dynamic study shows that
a single static topology leaves roughly a quarter of the recoverable losses on the
table once load and generation vary, and that a dynamic policy---especially one
that warm-starts each step from the previous solution---recovers most of it.

Several extensions follow naturally. The data-driven paradigm, supervised models and reinforcement-learning agents, should be evaluated on the same networks and the same NPF/loss metric, using the deterministic solvers benchmarked here to generate training labels; the open question is whether a learned policy generalizes across topologies and load patterns it was not trained on.

The dynamic study assumes switching is free, whereas switches have a finite mechanical life under load; incorporating a per-operation cost would turn the year-long policy into a constrained scheduling problem that trades recovered energy against asset cost. Finally, the mathematical formulation could be extended from the implemented mixed-integer quadratic program to the second-order cone relaxation of~\cite{taylor_convex_2012}, adding the voltage- and current-magnitude limits that the current model omits.

%% Bibliography ---------------------------------------------------------------
\bibliographystyle{unsrtnat}
\bibliography{bib1}

\end{document}